\documentclass[aps, prx,twocolumn,longbibliography,superscriptaddress,showpacs,amsmath,amssymb,floatfix]{revtex4-1}
\usepackage{amssymb}
\usepackage{graphicx}
\usepackage{amsmath}
\usepackage{color}
\usepackage{mathrsfs}
\usepackage{float}
\usepackage{indentfirst}
\usepackage{mathrsfs}
\usepackage{float}
\usepackage{indentfirst}
\usepackage{textcomp}
\usepackage{comment}
\usepackage{mathtools}
\usepackage{natbib,hyperref}
\usepackage{soul}

\begin{document}

\title{Absence of superconductivity in the pure two-dimensional Hubbard model}

\author{Mingpu Qin}
\thanks{These two authors contributed equally to this work.}
\affiliation{Key Laboratory of Artificial Structures and Quantum Control, School of Physics and Astronomy, Shanghai Jiao Tong University, Shanghai 200240, China}
\affiliation{Department of Physics, College of William and Mary, Williamsburg, Virginia 23187, USA}

\author{Chia-Min Chung}
\thanks{These two authors contributed equally to this work.}
\affiliation{Arnold Sommerfeld Center for Theoretical Physics, Ludwig-Maximilians-Universit\"{a}t M\"{u}nchen, 80333 Munich, Germany}
\affiliation{Munich Center for Quantum Science and Technology (MCQST), 80799 Munich, Germany}

\author{Hao Shi}
\affiliation{Center for Computational Quantum Physics, Flatiron Institute, New York, NY 10010, USA}

\author{Ettore Vitali}
\affiliation{Department of Physics, California State University Fresno, Fresno, California 93740, USA}
\affiliation{Department of Physics, College of William and Mary, Williamsburg, Virginia 23187, USA}

\author{Claudius Hubig}
\affiliation{Max-Planck-Institute for Quantum Optics, 85748 Garching, Germany}

\author{ Ulrich Schollw\"{o}ck}
\affiliation{Arnold Sommerfeld Center for Theoretical Physics, Ludwig-Maximilians-Universit\"{a}t M\"{u}nchen, 80333 Munich, Germany}
\affiliation{Munich Center for Quantum Science and Technology (MCQST), 80799 Munich, Germany}

\author{Steven R. White}
\affiliation{Department of Physics and Astronomy, University of California, Irvine, California 92697, USA}

\author{Shiwei Zhang}
\affiliation{Center for Computational Quantum Physics, Flatiron Institute, New York, NY 10010, USA}
\affiliation{Department of Physics, College of William and Mary, Williamsburg, Virginia 23187, USA} 

\collaboration{The Simons Collaboration on the Many-Electron Problem}

\begin{abstract}
%Using two complementary, state-of-the-art many-body computational methods, we 
We study the superconducting
pairing correlations in the ground state of the %pure 
doped Hubbard model -- in its original form without hopping beyond nearest
neighbor or other perturbing parameters -- in two dimensions
at intermediate to strong coupling and near optimal doping.
The nature of such
correlations has been a central question ever since the discovery of cuprate
high-temperature superconductors. Despite unprecedented effort and tremendous progress in understanding the
properties of this fundamental model, a definitive answer to whether the ground state is
superconducting in the parameter regime most relevant to cuprates has proved exceedingly difficult
to establish. In this work, we employ 
two complementary, state-of-the-art many-body computational methods,
constrained path (CP) auxiliary-field quantum Monte Carlo (AFQMC) and density
matrix renormalization group (DMRG) methods, deploying the most recent algorithmic advances in
each. Systematic and detailed comparisons between the two methods are performed. The DMRG is
extremely reliable on small width cylinders, where we use it to validate the AFQMC.  
The AFQMC is then used  to study 
wide systems as well as fully periodic systems, 
to establish that we have reached
the thermodynamic limit.
The ground state is found to be non-superconducting in the moderate to strong coupling regime
in the vicinity of optimal hole doping. 
\end{abstract}

\maketitle

\section{introduction}
\label{sec:intro}

Understanding high-temperature superconductivity in the cuprates \cite{Bednorz1986} has been
a long-standing mystery and one of the greatest challenges in theoretical
condensed matter physics \cite{RevModPhys.66.763}. Very early on the single-band two-dimensional (2D) Hubbard model \cite{Hubbard238}, along with its cousin,
the $t$-$J$ model, were argued to be the paradigmatic models for this problem \cite{ANDERSON1196,PhysRevB.37.3759}, 
and in many ways, this
suggestion has proven to be accurate.  Many of the properties of the cuprates seem to be reasonably 
well 
described -- or at least mirrored -- in the Hubbard model, such as antiferromagnetism \cite{PhysRevB.31.4403,PhysRevB.94.085140,PhysRevB.94.085103} and its 
abrupt disappearance upon doping, pairing and stripe formation, and pseudogap physics \cite{RevModPhys.84.1383}.  
Pairing, when it occurs, can be seen as
a consequence of a sort of frustration, between the hopping/kinetic energy of the holes and 
antiferromagnetic correlations, which are disrupted by the hopping.

Superconducting long-range order itself, however, is one of the most delicate 
properties in these systems.  Superconductivity appears to have a subtle competition and coexistence
with stripe formation \cite{Kivelson1998,RevModPhys.87.457,PhysRevB.97.075112}. In terms of the models, this means that accurate answers about a possible 
superconducting phase require simulations which are able to describe all of the possible phases
in an unbiased fashion, so that their competition can be resolved. One also needs a
systematic approach to the zero-temperature 
as well as the thermodynamic limit, particularly since stripes can introduce a new length
scale somewhat larger than the size of a pair. 
Numerous studies over the years have addressed pairing order in the
Hubbard model. 
They have often been driven by remarkable methodological advances, and have led to a great deal of insight in the 
physics of the model (see, for example, 
Refs.~\cite{doi:10.1143/JPSJ.57.2482,PhysRevB.38.931,PhysRevB.49.12318,PhysRevB.62.R9283,
	PhysRevLett.88.117002,PhysRevLett.95.237001,PhysRevLett.94.156404,PhysRevB.74.054513,
	PhysRevB.74.235117,PhysRevB.74.165122,doi:10.1143/JPSJ.76.113708,
	doi:10.1143/JPSJ.75.114706,doi:10.1143/JPSJ.73.545,PhysRevB.76.180504,PhysRevB.74.085104,	PhysRevB.85.081110,PhysRevB.86.241106,PhysRevLett.110.216405,PhysRevLett.115.116402,PhysRevB.90.115137,PhysRevB.93.035126,PhysRevB.98.205132,PhysRevLett.78.4486,PhysRevB.59.1706,PhysRevLett.81.1294,jiang2019ground}.
However, given the competing energy scales and intertwined states, it can reasonably be argued that none has satisfied these rigorous criteria
for establishing the nature of superconductivity in the physically relevant parameter regime.
Both positive 
and negative 
results have been found for $d$-wave pairing order, 
reflecting the extreme sensitivity of the ground state and low-lying excitations in the model, and the competition between
d-wave and other states \cite{MACHIDA1989192,PhysRevB.40.7391,doi:10.1143/JPSJ.59.1047,PhysRevB.39.9749,
	refId0,PhysRevLett.80.1272,PhysRevB.60.R753,PhysRevLett.88.117001,PhysRevLett.91.136403,
	PhysRevB.71.075108,PhysRevB.76.140505,PhysRevB.78.134530,PhysRevLett.104.116402,
	PhysRevB.84.041108,PhysRevLett.113.046402,PhysRevB.93.035126,PhysRevB.96.085103}.

The relation
between superconductivity and stripes 
%\cite{MACHIDA1989192,PhysRevB.40.7391,doi:10.1143/JPSJ.59.1047,PhysRevB.39.9749,
%	refId0,PhysRevLett.80.1272,PhysRevB.60.R753,PhysRevLett.88.117001,PhysRevLett.91.136403,
%	PhysRevB.71.075108,PhysRevB.76.140505,PhysRevB.78.134530,PhysRevLett.104.116402,
%	PhysRevB.84.041108,PhysRevLett.113.046402,PhysRevB.93.035126,PhysRevB.96.085103} 
	or other orders is also strongly affected by
modifications of the model, such as the next nearest neighbor hopping $t'$.
Given the existence of superconductivity in the cuprates with an apparent electronic
mechanism, it seems likely that some modification of the pure model exhibits superconductivity.
For example, recent studies on width-four cylinders---where DMRG can be pushed to resolve
the competing phases to high accuracy---found 
a non-superconducting filled-stripe state in the pure model, but quasi-long-range pairing
correlations with the addition of a $t'$ term, coexisting with half-filled stripes \cite{PhysRevB.95.125125,Jiang1424,jiang2019ground,ponsioen2019period}. 
While
superconductivity arising from the addition of a $t'$
is encouraging, one clearly needs to go beyond width four. 
 (Note that a width
four cylinder is equivalent to a stack of plaquettes, and there is no difference between a pair
on a plaquette and a half-filled stripe. Larger systems are needed to properly allow stripes
and superconductivity to compete or coexist.)
Hopping parameters $t'$ and third neighbor (diagonal) $t''$ have been predicted using 
electronic structure methods \cite{ANDERSEN19951573},
but even small differences in these parameters can alter the ground state phase
and it is difficult to establish whether additional terms, such as hopping mediated by a second hole, are important.
%or whether additional terms, such as hopping mediated by a second hole, are important.
It is also not clear
whether one needs to study a three-band model in order to connect directly with the cuprates.

Here, we choose 
to focus on the pure Hubbard model, with parameters $U$ and $t$ only. 
The existence or absence of superconducting order in this fundamental model at moderate to strong coupling is an outstanding theoretical question. 
This question has presented a 30-year challenge, magnified by the 
quest to understand  high-$T_c$ superconductivity.
An intense experimental effort is on-going with ultracold atoms in optical lattices to realize ``quantum simulations'' of this model \cite{PhysRevLett.116.175301,Brown379,article_nature_545_462,Koepsell_2019}.
The model has also served as a barometer for the capacity of the computational physics and chemistry community 
to perform reliable computations in interacting quantum systems. 
We study pairing correlations and superconductivity using two complementary methods, the density
matrix renormalization group (DMRG) and auxiliary field quantum Monte Carlo (AFQMC). Our work follows up
on a previous study involving four different methods which determined that the ground state
of the Hubbard model has stripe order at $1/8$ doping \cite{Zheng1155}. Although stripes may tend to compete with superconductivity \cite{1367-2630-15-9-093005},
it may be possible for them to coexist \cite{PhysRevB.78.174529,PhysRevB.83.104506,PhysRevB.97.075112,MIYAZAKI20021403,PhysRevB.98.205132,2017arXiv171204297W}.

The constrained path AFQMC method \cite{PhysRevB.55.7464,PhysRevB.94.235119} we use treats the fermion sign problem approximately,
so 
validation is important. Here we use DMRG \cite{PhysRevLett.69.2863} on width four and six cylinders to validate
an approach to predict pairing orders in AFQMC. The DMRG calculations involve multiple
independent DMRG programs pushed to the limit of current capabilities. We find excellent
agreement between the DMRG and AFQMC.  The AFQMC does not have DMRG's width restrictions,
and we then use the AFQMC to study systems %as large as 
of over $250$ lattice sites, %\C{$32 \times 8$}, %$16\times 16$,
including periodic
boundary conditions. 
In the AFQMC calculations, we devise new techniques to probe the superconducting order, both through a linear response measure of the 
order parameter, and through the use of a BCS trial wave function to directly measure the pairing correlation function.
These simulations allow
us to 
conclude
that only short-range pairing occurs in the regime of interest
($U/t$ around 
$6$-$8$ 
and dopings $0.1 < h < 0.2$),
and the system is not superconducting. 

In the small $U/t$ limit, controlled results from perturbation theory have shown that the Hubbard model has a superconducting ground state 
\cite{PhysRevB.46.11163,PhysRevB.48.1097,0295-5075-56-4-563,PhysRevB.81.224505}.
Diagrammatic Monte Carlo studies \cite{Deng_2015} indicate that a BCS superconducting state of $d$-wave symmetry can emerge 
at weak coupling ($U/t< 4$)
for doping $h\ge \sim 0.3$. 
Given the sensitive and delicate nature of the ground state of the model, and, in particular, given that stripe formation is 
believed not to occur at weak coupling\cite{PhysRevB.81.224505},
it is a very interesting question how this part of the phase diagram connects with the other parameter regimes.
We emphasize that our work does not imply a general statement that there is no superconducting order anywhere in the pure 
Hubbard model.
Rather our focus is on the nature of the pairing order in the pure Hubbard model in the physically important parameter regime as a model for cuprate
 superconductors. 
 
The rest of this paper is organized as follows. 
Section~\ref{sec:approach} discusses the two different methods we employ, and two different ways in each to probe pairing and superconducting order.
Our results are presented in Sec.~\ref{sec:results}: first a general scan of the doping dependence of the superconducting order at $U=8$, 
then a detailed study of the case of $U=8$ and $h=1/8$, followed by an analysis of the relation between pairing and 
stripe order, and then the dependence on the interaction strength. We conclude in Sec.~\ref{sec:summary}.
 Further technical details as well as additional results are included in the Appendix.
 
\section{Approach}
\label{sec:approach}

We study the pure Hubbard Hamiltonian with nearest-neighbor hopping and on-site interaction:
\begin{equation}
	\hat{H}=-t\sum\limits _{\langle ij\rangle\sigma}\hat{c}_{i\sigma}^{\dagger}\hat{c}_{j\sigma}+U\sum\limits _{i}\hat{n}_{i\uparrow}\hat{n}_{i\downarrow}-\mu\sum_{i\sigma}\hat{n}_{i\sigma},
	\label{eqn:H_pairing}
\end{equation}
where $\hat{c}_{i\sigma}$ is the fermionic annihilation operator, $\sigma$ denotes spin ($=\uparrow$ or $\downarrow$),
$\hat{n}_{i\sigma}=\hat{c}^\dagger_{i\sigma} \hat{c}_{i\sigma}$ is the particle number operator
on site $i$, and
$\langle ij\rangle$ denotes nearest neighbor sites.
We study rectangular lattices of size
$N = L_x \times L_y$,
typically with periodic boundary conditions (PBC) along
the $y$ direction and open boundary conditions along the $x$ direction (i.e., cylinder geometry).
We vary the aspect ratios of the cylinders (e.g., $32 \times 8$, $24 \times 14$) to ensure that 
the rectangular cells do not impact our results \cite{PhysRevLett.99.127004}. Finite-size extrapolations are performed. Additionally,
complementary calculations 
are performed with PBCs along both directions. 
We set $t$ as the energy unit, i.e., $t = 1$. 

We denote the number of electrons in the simulation cell by $N_e$, with $N_e=N_\uparrow + N_\downarrow$. The electron density or 
 filling factor is $n = N_e / N$, and the hole doping level is then $h = 1 - n$. 
 These quantities are specified in an average sense, as $N_e$ is controlled by the chemical potential $\mu$ and will fluctuate
 in most of our calculations.

\subsection{Two complementary methods}
\label{subsec:2methods}

In this work we employ two state-of-the-art methods, constrained path (CP) %auxiliary field quantum Monte Carlo (AFQMC) and the density matrix renormalization group (DMRG). 
AFQMC and DMRG.
These methods are representative of the leading edge of computational capabilities for interacting quantum many-fermion systems. 
They involve very different approximations in
obtaining ground-state properties in the thermodynamic limit.
To quantify the
CP error in AFQMC,
we benchmark the results in finite systems of narrow cylinders, where DMRG is highly accurate.  
The AFQMC does not have size or boundary condition restrictions and can reliably approach the thermodynamic limit.
 The systematic, detailed, and
complementary use of these leading 
computational techniques is a unique and distinguishing feature of the present study. 
The excellent agreement between the two methods allow us to draw conclusions with confidence.

\subsubsection{Constrained-path auxiliary field quantum Monte Carlo}

In AFQMC,
the interaction part of the Hamiltonian is re-cast into a summation (or an integral) of
non-interacting terms through a Hubbard-Stratonovich transformation. As a result, physical quantities are represented as a path integral in 
many-dimensional 
auxiliary field space. 
The high-dimensional summation or integral can be evaluated with Monte Carlo techniques \cite{PhysRevB.31.4403}. However, with few exceptions,
a minus sign problem is present
\cite{PhysRevB.41.9301} which causes an exponential growth of the statistical errors with system size.
The CP approach overcomes this difficulty by imposing a boundary condition in  auxiliary field space, which is derived from an exact property of the path integral \cite{PhysRevB.55.7464} but whose practical implementation involves a trial wave function.
The use of CP introduces a systematic error, which can be improved with better trial wave functions.
 Usually simple wave functions such as the Hartree-Fock solution have been used as trial
wave-functions and previous results \cite{PhysRevB.78.165101}
show the systematic error is typically small. Recently we introduced an approach \cite{PhysRevB.94.235119} to optimize the trial
wave-function self-consistently, further reducing the systematic error. 
As mentioned, a key feature of this work is the combined use of 
CP-AFQMC with DMRG, which allows us to systematically gauge the accuracy of CP in cylindrical 
systems.

Ground-state AFQMC is typically formulated in a  sector of the Hilbert space with fixed number of particles, $N_e$, and fixed $S^z$ (although 
a corresponding approach in Hartree-Fock-Bogoliubov space exists \cite{PhysRevB.95.045144}). 
 Our computation of the pair-pair correlation function is done in this manner, by separate AFQMC calculations 
 on the original Hubbard Hamiltonian in 
  Eq.~(\ref{eqn:H_pairing}), 
  using back-propagation \cite{PhysRevB.55.7464} and BCS trial wave functions \cite{PhysRevA.100.023621}.
In this work the order parameter  is computed in AFQMC after a particle-hole transformation has been applied to
 Eq.~(\ref{eqn:H_pairing}), which results in a modified Hamiltonian that conserves the total particle number \cite{PhysRevB.59.1706} 
 but breaks total $S^z$ (further details in Appendix). 
 As described in the next section, in this formulation the order parameter can be computed from total energy 
 calculations, which leads to very accurate results.

\subsubsection{Density matrix renormalization group}
DMRG is a variational method\cite{PhysRevB.48.10345,PhysRevLett.69.2863} which can be understood in the language of matrix product states (MPS)\cite{SCHOLLWOCK201196}.
The MPS matrix dimensions, or the so-called bond dimensions, indicate the number of states kept in the reduced Hilbert space, and play central roles in the approximation.
A general many-body state can be represented by an MPS with exponential growth of the bond dimension from the edges.
In practice one restricts the maximum value of the bond dimension, thus limiting the maximum entanglement allowed in the variational state. 
Ground states of local Hamiltonians of physical interest generally have low entanglement.
DMRG minimizes the energy in this low-entanglement Hilbert space.
The accuracy of DMRG can be systematically improved by increasing the bond dimension.
Although DMRG is naturally formulated and most powerful for one-dimensional systems, 
it is now widely applied to 2D systems\cite{doi:10.1146/annurev-conmatphys-020911-125018}, and remains one of the most-accurate numerical methods in 2D.

In this work, we employ two DMRG schemes with different conserved quantum numbers and using different update schemes.
The first scheme conserves only the $S^z_\mathrm{tot}$ 
with $U(1)$ symmetry, and uses the two-site update in the optimization.
This scheme is used when a pairing field is applied to the system, breaking the particle number conservation.
In such systems the particle numbers are controlled by the chemical potential.
This scheme efficiently enables fluctuations between different quantum numbers in the optimization and is less likely to be stuck in a local minimum.
The truncation errors in this scheme are in the order of $10^{-7}$ (smaller doping) to $10^{-5}$ (larger doping).
The second scheme \cite{hubig:_syten_toolk} conserves both the $U(1)$ total particle number and $SU(2)$ spin symmetries, and uses the single-site update\cite{hubig17:_symmet_protec_tensor_networ,PhysRevB.91.155115}.
The single-site update is faster than the two-site update and thus allows us to achieve large bond dimension.
This scheme is used for systems without pairing fields, which thus conserve total particle number.
Since the truncation error is ill-defined in the single-site update, we use the two-site energy variance in the standard extrapolations~\cite{PhysRevB.97.045125}.
The number of states kept in these systems is up to $30000$ $SU(2)$ states, which corresponds to $\sim 90000$ $U(1)$ states, providing the best accuracy attained to date to our knowledge.

\subsection{Two different ways to characterize superconducting correlation}
\label{subsec:2measures}

\begin{figure*}[t]
	\includegraphics[width=58mm]{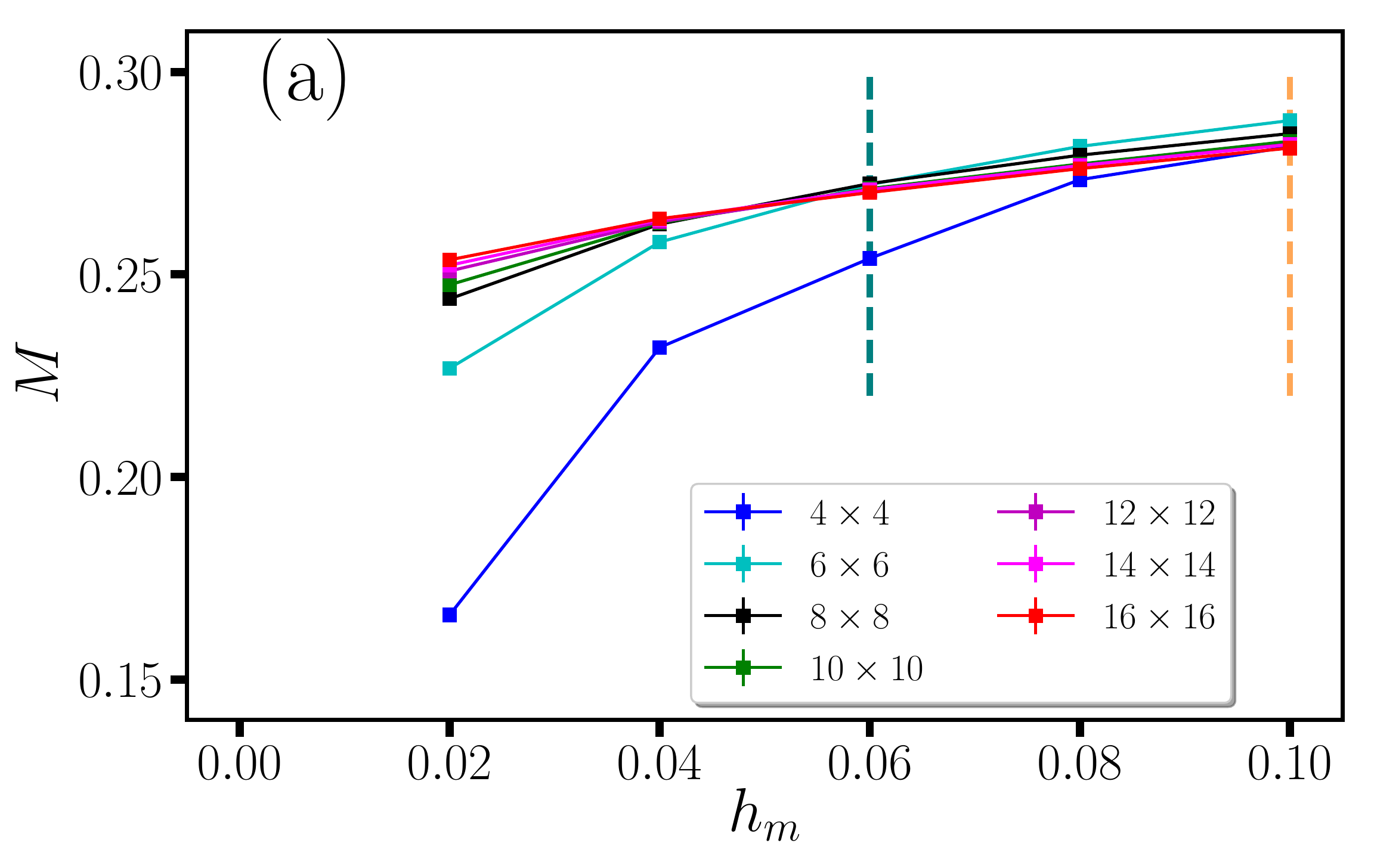}
	\includegraphics[width=58mm]{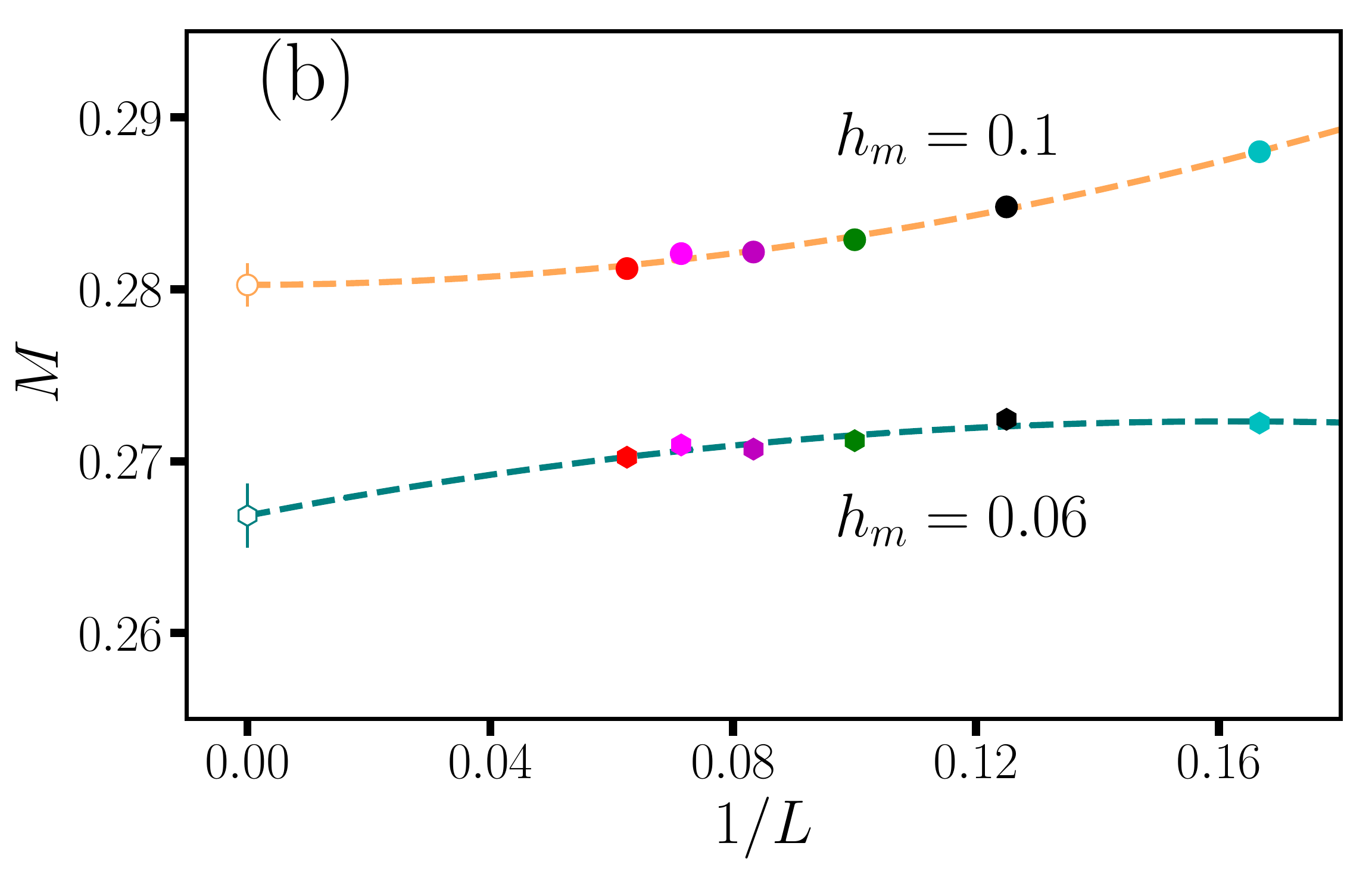}
	\includegraphics[width=58mm]{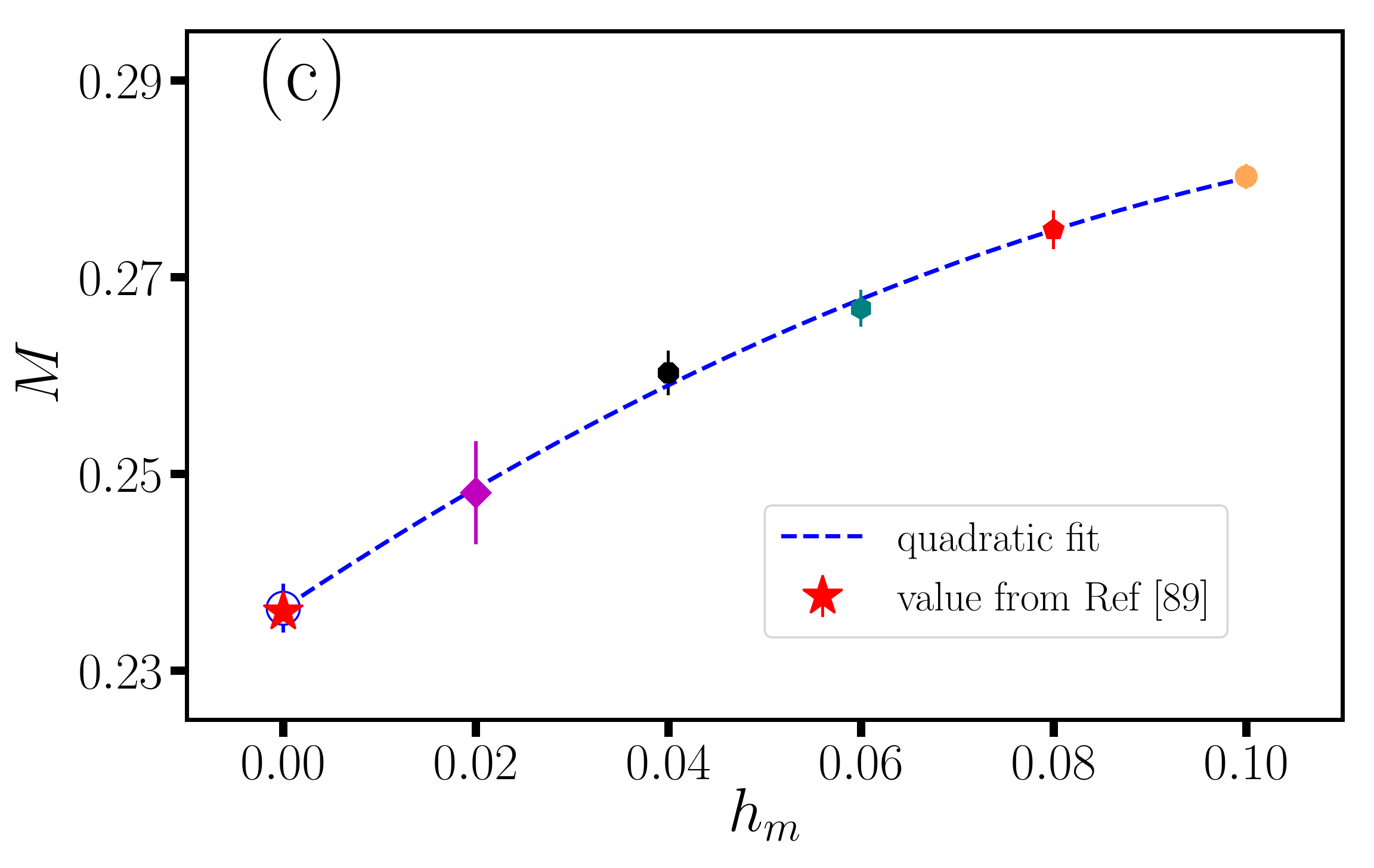}
	\caption{Illustration of the approach to compute order parameters.
The AFM order parameter at half-filling ($U = 4$)
is computed by applying staggered magnetic fields to the periodic supercell.
		(a) The computed magnetic order parameter $M$ is shown as a function of the pinning field strength $h_m$ for different system sizes.
		(b) A quadratic fit is performed at each $h_m$ to 
		extrapolate $M(h_m)$ to the TDL. The procedure is shown for two values 
		 $h_m = 0.06$ and $0.1$, as marked by the vertical dash lines in (a).
		 (c) 
		An extrapolation of $M_\infty(h_m)$ to the $h_m\rightarrow 0$ limit is then performed, using again a quadratic fit.
		 The resulting $M_\infty(0)$ is shown by the open symbol. 
		The value is in excellent agreement with the order parameter determined from direct computation of spin-spin correlation functions 
		\cite{PhysRevX.5.041041}, shown by the red star.
	}
	\label{mag_size_scaling}
\end{figure*}

To study the superconducting properties in the ground state, we use two different probes: pair-pair correlation functions and the pairing order parameter.
 These are both defined in terms of the pairing operator of
 a pair of nearest-neighbor sites, $i$ and $j$:
 \begin{equation}
	\hat{\Delta}_{ij}\equiv\frac{(\hat{c}_{i\uparrow}\hat{c}_{j\downarrow}-\hat{c}_{i\downarrow}\hat{c}_{j\uparrow})}{\sqrt{2}}\,.
	\label{eqn:Delta_def}
\end{equation}
 We will compute the  
 pair-pair correlation function 
\begin{equation}
	P_{i'j',ij}=\langle \hat{\Delta}_{i'j'}^{\dagger} \hat{\Delta}_{ij} \rangle\,,
	\label{eqn:pair-corr-def}
\end{equation} 
and the pairing order parameter
\begin{equation}
	\Delta_{i,j} = \langle (\hat{\Delta}_{ij} + \hat{\Delta}_{ij}^{\dagger})/2\rangle\,,
	\label{eqn:order-param-def}
\end{equation} 
where $\langle \cdots\rangle$ denotes expectation with respect to the many-body ground state.

  The pair-pair correlation function in Eq.~(\ref{eqn:pair-corr-def})
  can be obtained directly in a calculation working in a sector with fixed particle numbers.  
  From it the $d$-wave pairing correlation function, $P^d(i-i')$,
 can be constructed as a function of pair separation $(i'-i)$, by considering 
all $j$ in $\langle ij\rangle$ and all $j'$ in  $\langle i'j'\rangle$, following the sign convention for $d$-
wave as we specify next.

The pairing order parameter in Eq.~(\ref{eqn:order-param-def}), on the other hand, requires a different approach.
We add a term in the Hamiltonian describing SC pairing fields \cite{PhysRevLett.99.127004,PhysRevB.79.220504} applied to the system:
\begin{equation}
	\hat{H}_p = -\sum_{\langle i,j\rangle}h_{p}^{ij}\frac{\hat{\Delta}_{ij}+\hat{\Delta}_{ij}^{\dagger}}{2}\,,
	\label{eqn:Hp}
\end{equation}
where the amplitude of $h_{p}^{ij}$ is given by 
the parameter $h_p$, 
and the sign of $h_{p}^{ij}$ 
is positive if  the bond $(i,j)$ is vertical (along $\hat y$-direction) and negative otherwise (along $\hat x$-direction), 
in order to probe  
pairing order of structure $d_{x^2-y^2}$ \cite{PhysRevB.39.839,Scalapino1999}.

In AFQMC, we can obtain the superconducting pairing order parameter $\Delta$ from total energy calculations, using
the Hellmann-Feynman theorem:
\begin{equation}
\Delta(h_{p})\equiv \Big \langle \frac{d(\hat{H}+\hat{H}_p)}{dh_p}\Big\rangle_{|\Psi_0(h_p)\rangle}=
\frac{dE(h_p)}{dh_p}\Big\vert_{h_p}\,,
	\label{eqn:HF-order}
\end{equation}
where $|\Psi_0(h_p)\rangle$ and $E(h_p)$ are the ground-state wave function and energy of the Hamiltonian $(\hat{H}+\hat{H}_p)$.
We compute the derivative in Eq.~(\ref{eqn:HF-order}) 
by finite difference $\Delta(h_{p})=\frac{E(h_{p}-\delta)-E(h_{p}+\delta)}{2\delta} + \mathcal{O}(\delta^2)$,
where $\delta$ is chosen to be sufficiently small to ensure that the error is smaller than our statistical error bar or targeted resolution. 
As $h_p\rightarrow 0$, the order parameter in the unperturbed ground state is obtained. 
This approach allows us to directly compute 
the pairing order parameter in AFQMC, which had not been possible before.

We next use an example to illustrate 
the above approach to compute order parameters.
We consider the antiferromagnetic (AFM)  Neel  order at half-filling.
A staggered inducing field is applied to the periodic supercell of size $L_x= L_y=L$, with magnitude $h_m$ and alternating signs on the two sublattices. 
Because of the absence of the sign problem at half-filling, no constraint is needed in the AFQMC 
calculation, and the results are exact numerically.
In Fig.~\ref{mag_size_scaling}(a), we show the computed staggered AFM 
order parameter $M_L(h_m)$ as a function of
the applied field strength $h_m$ for different lattice sizes. Extrapolation to the thermodynamic limit (TDL) is then performed at each fixed
 $h_m$, as illustrated in panel (b).
 The resulting TDL values are plotted in panel (c) versus $h_m$, and extrapolated to the $h_m\rightarrow 0$ limit to obtain the order parameter.
  The result of $0.236(3)$ is in excellent agreement with the previous result of $0.236(1)$ 
  computed from spin-spin correlation functions  \cite{PhysRevX.5.041041}. 
 This test provides a validation of our approach for computing superconducting order parameters, which follows identical procedures.
 (We note that the staggered AFM magnetization in the 
 repulsive Hubbard model at half-filling can be mapped to the $s$-wave on-site pairing order parameter in the attractive Hubbard model,
 through a partial particle hole transformation as discussed in the Appendix.)

In the following two subsections, we show benchmark results on the two ways to compute the
pairing order, respectively.
Careful and detailed comparisons are made between AFQMC and DMRG, first 
for computing the
pairing order parameter and then for pair-pair correlation functions,
by using cylindrical geometries.
Our results applying these approaches to address the physical properties of the Hubbard model are presented in Sec.~\ref{sec:results}.

\subsubsection{Pairing order parameter}

\begin{figure}[t]
	\includegraphics[width=80mm]{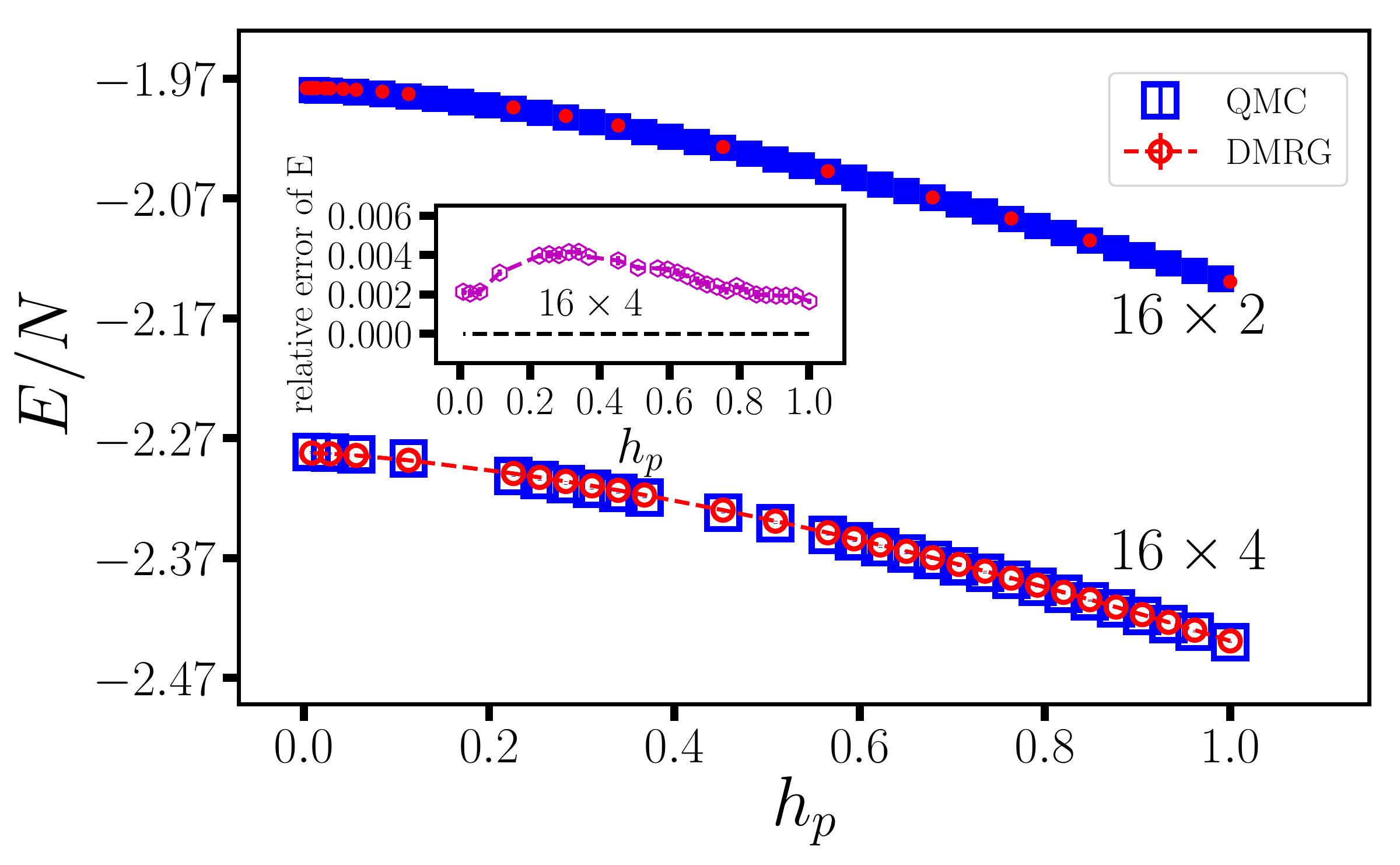}
	\includegraphics[width=80mm]{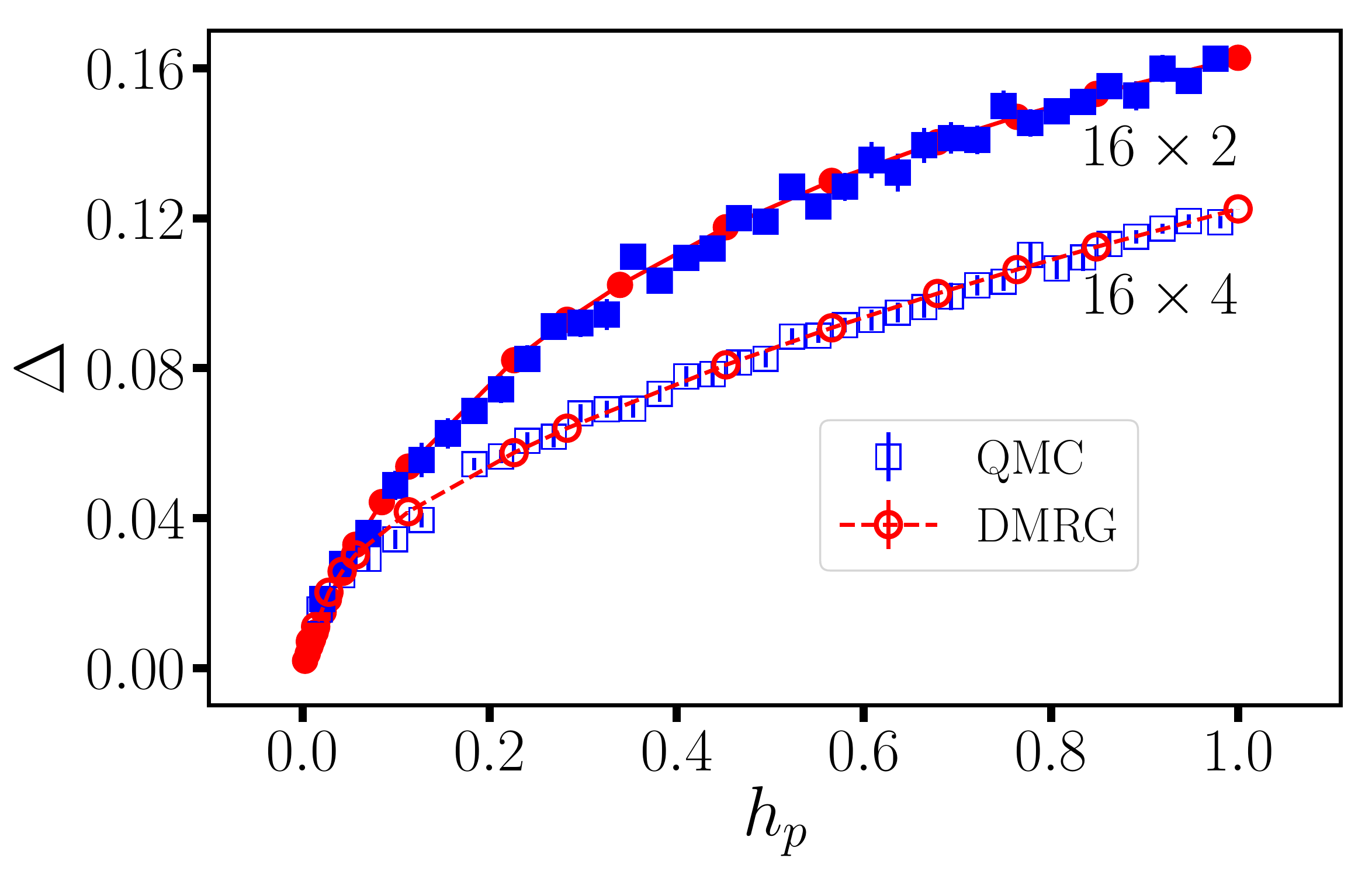}
	\caption{
		%\C{({\em Added  data between $h_p = 0.2$ to $h_p = 0.4$ in the upper panel.})}
		Upper: Comparison of the ground-state energies computed from AFQMC (blue) and DMRG (red),
        as a function of the applied pairing field
		strength $h_p$, at $U=8$. Two cylindrical systems are shown,  $16 \times 2$ and 
		$16 \times 4$, with chemical potential held fixed in each so that the doping is $1/8$ when $h_p \rightarrow 0$. 
		The inset shows the difference of the energy computed from AFQMC with respect to DMRG.
		Lower: Comparison of the computed SC pairing order parameter 
		for the same systems. 
		}
	\label{E_pair_4_16}
\end{figure} 

\begin{figure}[t]
	\includegraphics[width=84mm]{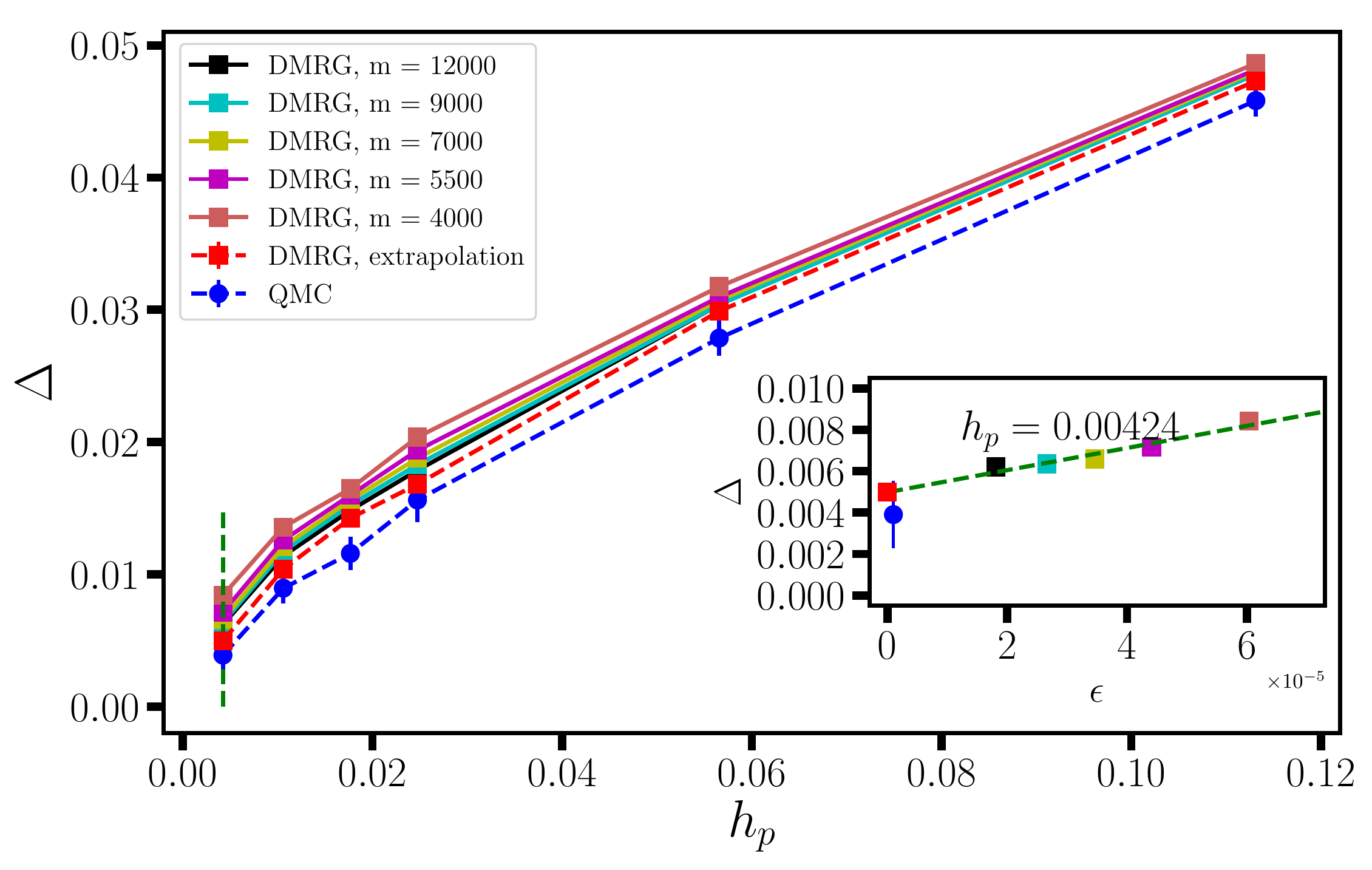}
	\caption{ 
		%\C{({\em Added more data}.)}
		Comparison of the computed pairing order parameter at $U=4$, as a function of the applied pairing field
		strength $h_p$. The system is a $24\times 4$ cylinder, at $h \simeq 1/6$. DMRG, with $U(1)$ symmetry and two-site update, results
        with different bond dimensions are shown. The results after extrapolation to zero truncation error are also plotted. 
		The inset illustrates the extrapolation (with respect to the truncation error) at $h_p =0.003 \times \sqrt{2} \approx 0.00424$. 
		The red dot represent
		the extrapolated value from DMRG, while the blue dot with error bar  is the AFQMC result.    
	}
	\label{4_24_1_6_U4_benchmark}
\end{figure}

The ground-state energies of $16 \times 2$ and $16 \times 4$  systems computed from AFQMC and DMRG are shown  in the top %left 
panel of Fig.~\ref{E_pair_4_16}, as a function of the applied pairing field
strength $h_p$. Uniform
``$d$-wave" pairing fields are applied to the entire system. 
A fixed value of $\mu $ is used which gives a doping of $1/8$ at $h_p = 0$
($\mu= 1.75$ for $16 \times 4$ and $\mu= 1.55$ for $16 \times 2$).
In AFQMC the trial wave-functions are optimized self-consistently  by coupling to natural orbitals \cite{PhysRevB.94.235119}.
(For small $h_p$, the resulting trial wave function is the same as the non-interacting wave function.)
The inset shows  difference between the energies computed from  AFQMC and DMRG.
The relative error of  the AFQMC energy is less than $0.5 \%$ for all $h_p$ in Fig.~\ref{E_pair_4_16} which means the CP  error
is very small.

In the bottom 
panel of Fig.~\ref{E_pair_4_16}, we plot the pairing order parameters
from AFQMC and DMRG, for the  same
system.
In DMRG the order parameter is directly computed as a ground-state expectation value for each $h_p$,
while in AFQMC it is computed with the approach involving Hellman-Feynman theorem described above. 
Agreement between the two methods is excellent throughout the entire range.
The general behavior of the order parameter is similar to that of the AFM order in Fig.~\ref{mag_size_scaling} for small supercell sizes. 
The pairing order parameter approaches $0$ linearly as $h_p\rightarrow 0$, 
which is reasonable as
spontaneous symmetry breaking can only occur in the TDL. 
At small $h_p$ a rapid drop is seen in  $\Delta$, deviating from the trend at larger $h_p$. 
The behavior is also manifested in the energy results as we show in the appendix:
a fit of the energies at $h_p>h_p^{\rm th}$ (where $h_p^{\rm th}$ is a threshold whose precise value does not affect the result), 
gives a $E_{\rm SC}(h_p=0)$ which lies above the true ground state energy of the system.

We also show the comparison for the pairing order parameter at $U = 4$ and $h = 1/6$ in a $24\times 4$ cylinder, in Fig.~\ref{4_24_1_6_U4_benchmark}.
Non-interacting trial wave-functions are used in the calculation. 
Note that the pinning field range here is much smaller than in  Fig.~\ref{E_pair_4_16},
focusing on the weak fields and a very fine scale of the
pairing order parameter for comparison.
With the lower value of $U$, this system requires larger bond dimensions in DMRG to converge, and we illustrate the extrapolation with  truncation error. 
Good agreement is seen in the order parameters computed from AFMQC and the extrapolated results from DMRG.

\subsubsection{Pair-pair correlation function}
The pair-pair correlation function is computed with fixed number of particles (canonical ensemble).
Results from CP-AFQMC  have been obtained earlier in supercells with PBC using free-electron trial wave functions \cite{PhysRevLett.78.4486} and 
also using a BCS type of trial wave function after a particle-hole transformation \cite{PhysRevB.59.1706}. 
Here we employ a more direct and general approach to apply projected BCS trial wave functions \cite{PhysRevA.100.023621}, and are able to access much larger 
systems because of algorithmic improvements and especially increased computing power.  More unique to this work is the detailed and direct comparison 
with DMRG to quantify the accuracy.

\begin{figure}[t]
	\includegraphics[width=85mm]{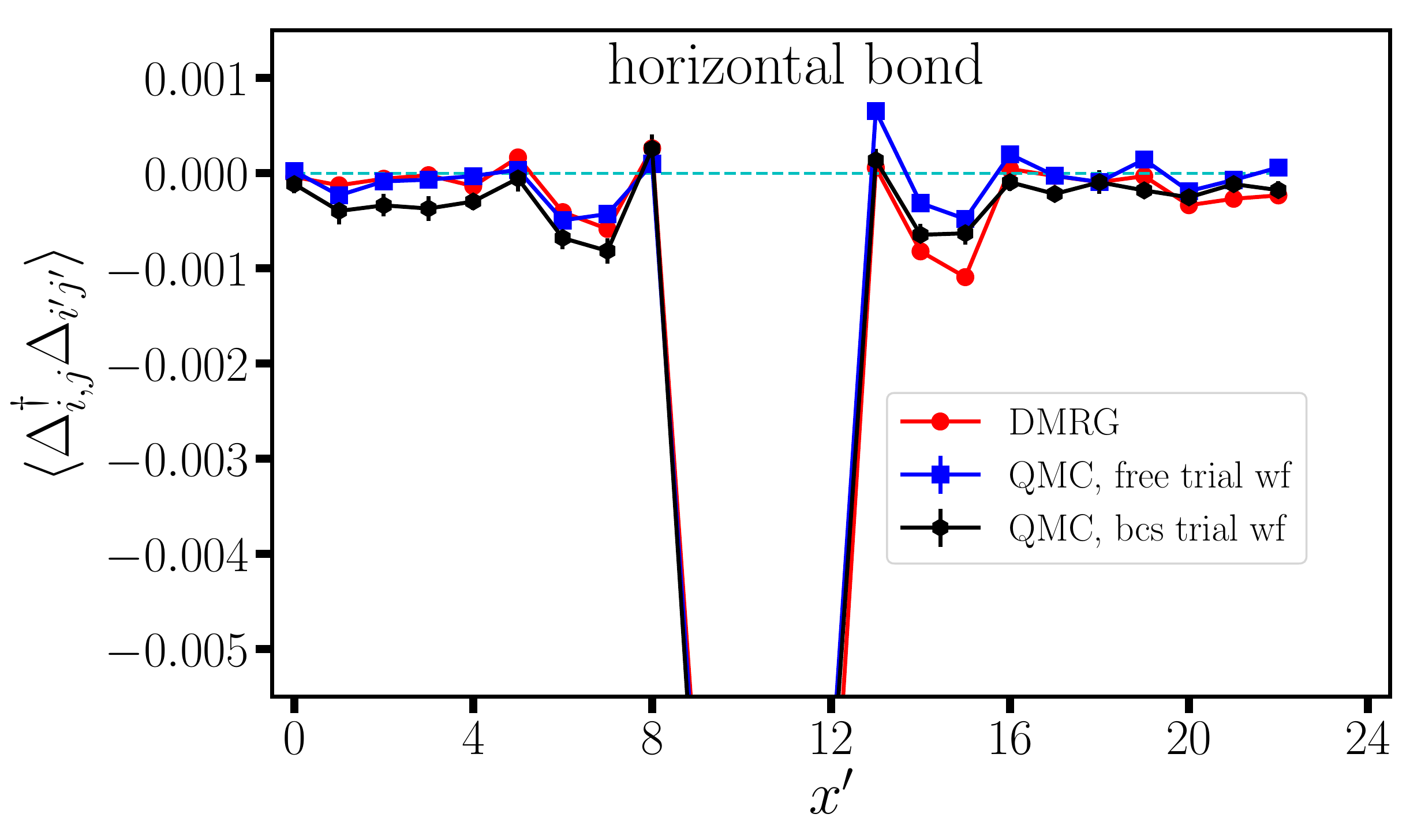}
	\includegraphics[width=85mm]{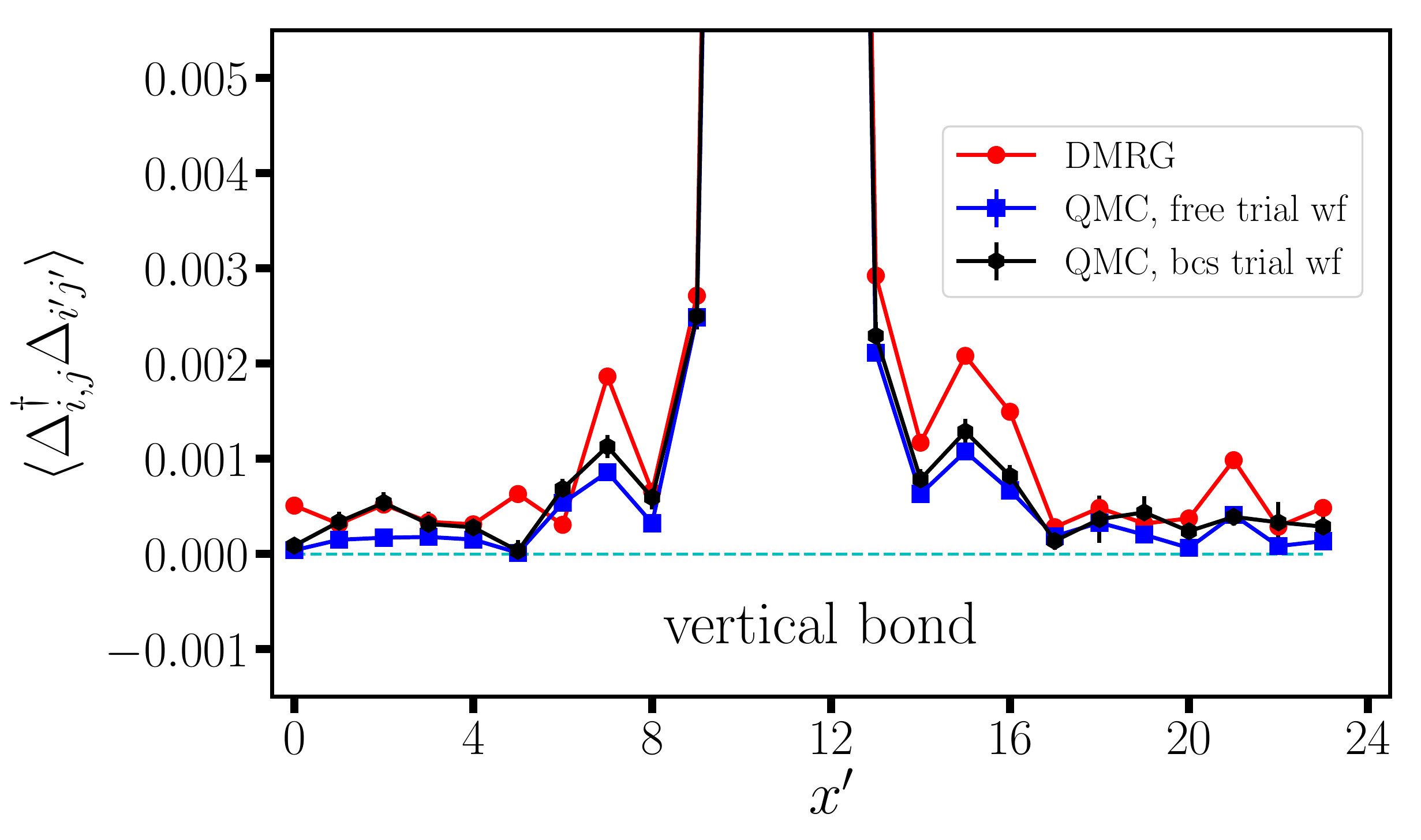}
	\caption{Comparison between DMRG and AFQMC of the pair-pair correlation function for a $24 \times 4$ cylinder, with $U = 4$ at $1/8$ doping.
		The reference bond is [(12,1), (12,2)]. Vertical and horizontal bonds are
		 along $y$ and $x$ direction respectively.
        %The number of states kept in DMRG is $15000$, with corresponding truncation error $\approx 4\times 10^{-6}$.
	}
	\label{4_24_U4_com_corr}
\end{figure}

In Fig.~\ref{4_24_U4_com_corr}, we show a comparison of the pair-pair correlation function for $24 \times 4$, at doping of $1/8$, with $U = 4$ in a cylindrical 
geometry
between DMRG and AFQMC. 
The pair-pair correlations have a much smaller signal (roughly $\Delta^2$), as can be seen even in these small system sizes.  Agreement is reasonable but the accuracy does not reach the level seen with the order parameter calculations. Hence the order parameter
will be the primary tool on which we rely to accurately determine the nature of 
superconducting orders. 
This is why the development in this work of direct computation of the order parameter
is crucial.

\subsection{Competition between pairing and stripes: small cylinder finite size effects}

A central question to this work is how stripe order and pairing order compete or
cooperate.  The interaction between the two types of order plays out in subtle
ways on small diameter cylinders. In this section we discuss this interplay 
in a general, intuitive way, %how this coexistence of orders can play out, 
as a guide for the subsequent finite
size scaling analysis.

Let us think about how one can have a striped state which also has pairing order. 
One would expect that two different requirements should be associated with such a state.
The first is that an individual stripe would have local pairing -- that one can think
of a stripe as being made up of pairs. The second is that, in order to have long-range
phase coherence, particularly between stripes, there should be substantial pair
tunneling between stripes, associated with a density of off-stripe pairs. One might think
of this as like a vapor pressure of pairs outside the ``liquid" of stripes. Stripes which
bind their pairs too tightly would have weak or nonexistent long-range pairing order.
Numerical approaches, in order to probe the 2D thermodynamic limit, must connect to
these two requirements.

Previous work on stripes has touched on the local pairing question.  Filled stripes were
first obtained theoretically in Hartree Fock calculations, which are based on a 
single-particle mean field approximation, without any notion of pairing.  If filled stripes have
pairing, it would have to be as a subtle modification of the non-paired mean field state. 
In contrast, there is evidence that partially filled stripes have local pairing structure.
For example, in DMRG simulations of the %t-J \Y{and Hubbard} model on cylinders, 
$t$-$J$ and Hubbard models on cylinders, 
it was noticed that the
ring-shaped stripes circling the cylinder strongly favor an even number of holes. For example,
on a width 6 cylinder, one finds stripes with either four or six holes, not three or five.
Another example is shown in the Appendix~\ref{Sec:supp_doping}.
In DMRG simulations without particle-number conservation, the hole number in a stripe is always even as the
  chemical potential is varied.)

To study whether pairs can leave their stripes, it is essential that the pair and the
stripe are distinct.  On a two leg ladder, there are only pairs, so one cannot address
this question. On a width four cylinder, a half-filled stripe is a pair, so one cannot
expect to probe the 2D physics very well on this system. A 4-hole filled stripe on
a width four cylinder  would
allow probing of pairs leaving the stripe, but because the stripe is filled, it may
not support pairs within the stripe.  Thus, the smallest cylinder which can address both
key questions has width six, where one can have a four hole stripe circling the cylinder.

Note that width four cylinders have another complication, unrelated to the thermodynamic
limit.  As discussed for the case of the $t$-$J$ model \cite{PhysRevB.95.155116}, there are two
very distinct forms of $d$-wave pairing on a width four cylinder. One is the usual type,
living on the surface of the cylinder.  The unusual type forms pairs circling the
cylinder, for which it is useful to think of the cylinder as a stack of plaquettes.  
It has been known for some time that a single plaquette nicely fits a $d$-wave pair.
This state seems especially 1-D-like.  Note that next nearest neighbor hopping $t^\prime$ connects sites
within pairs for the surface pairing state, but not for the plaquette state.

\section{Results}
\label{sec:results}

This section contains the following four parts. We first  scan the pairing susceptibility versus doping $h$ at a representative interaction strength of $U=8$.
We then carry out a detailed and systematic study of the pairing properties at $1/8$ and $U=8$ in Sec.~\ref{ssec:1eigthdoping}. 
This is followed by an examination of the relation between stripe and SC orders in 
Sec.~\ref{ssec:stripevSC}, and then an investigation of the dependence on $U$ in Sec.~\ref{ssec:U4}.

\subsection{$d$-wave pairing susceptibility versus doping}
\label{ssec:doping-dep}

\begin{figure}[h!]
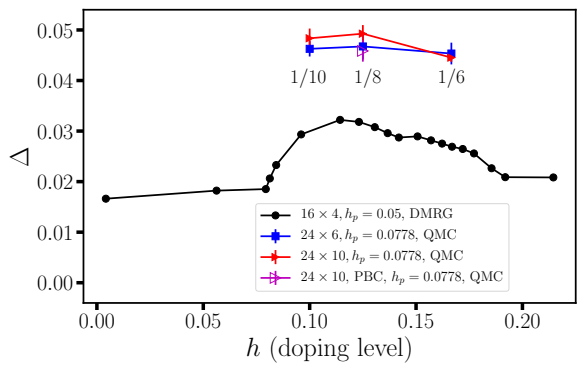

       \includegraphics[width=0.9\columnwidth]{{{pair_diff_filling}}}
        \caption{ 
        	%\C{(\em{Added PBC data.})}
        SC order parameters as a function of doping level, with $U=8$. Results from three systems are shown, each at a modest value of the applied $d$-wave pairing field (with 
        strength $h_p$ indicated in the legend).  
        The doping level or particle density is controlled by varying the chemical potential.
        }
        \label{fig:pair_vs_n}
\end{figure}

We first probe the SC response as a function of electron density, by computing the pairing order parameter in the presence of a 
$d$-wave  pairing field, which is applied to the entire system.
We choose the pairing field amplitude around $h_p=0.05$, %the pairing field amplitude $h_p$ is chosen to be  of modest strength,
which induces a sizable SC order but does not drive the system 
far away from its ground state (see Fig.~\ref{fig:compare_hole} in Appendix.~\ref{Sec:supp_doping}).
The electron density $n$ is controlled by the chemical potential $\mu$.
For the $16\times 4$ cylinder, the $\mu$ value was varied in the range of $1.4$ to $2.0$ which yielded an electron density from $\sim 0.79$ to $1$.
Fig.~\ref{fig:pair_vs_n} shows the SC pairing order parameter as a function of density, or doping level.
It can be seen from the  $16\times 4$ scan that the SC order has stronger response between $n\sim 0.81$ and $0.92$, with the maximum close to $n=0.885$.
Results in wider systems remain consistent, with the SC order showing slow variations in the vicinity of $h=1/8$ doping.
At density near the maximum SC order, 
the system displays charge and spin orders consistent with the ground state at $1/8$ doping ($n=0.875$), namely a stripe order
\cite{Zheng1155}.
This indicates that the system shows a SC  order in response to the applied pairing field but remains in a similar ground state as the one when the pairing field is 
absent.

We have also investigated the doping dependence of the SC response in a $64\times 4$ system using a different and complementary approach to the one in Fig.~\ref{fig:pair_vs_n}. A linearly varying chemical potential, $\mu(x)$, is applied along the cylinder, and the SC order and local density are computed
without pairing field by allowing particle numbers to fluctuate in the DMRG calculation. The dependence of the SC order as a function of local density is found to be consistent
with that in Fig.~\ref{fig:pair_vs_n}, as shown in the Appendix.

The fact that $1/8$ doping is near the maximum response of the SC order for $U=8$, and that the SC order shows rather weak dependence on the precise density, 
leads us to focus on the system of $h=1/8$, $U=8$, for which there is also detailed data on the spin and charge order, as well as ground-state energy, to compare with. The interaction strength of $U=8$ is chosen as representative of the physically relevant regime.  
The results are presented in the next section.

\subsection{Absence of long-range $d$-wave
pairing order at $U = 8$ and $h = 1/8$}
\label{ssec:1eigthdoping}

\begin{figure}[h!]
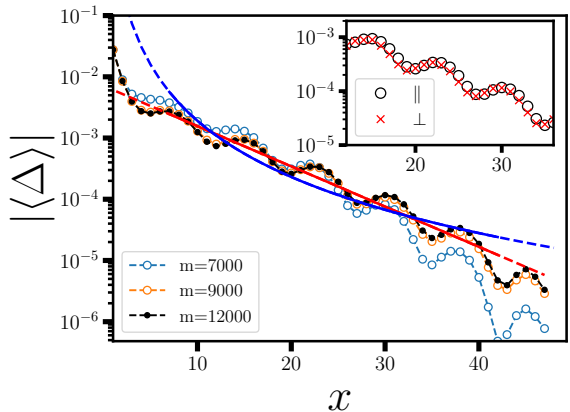
	
	\includegraphics[width=0.9\columnwidth]{{{huben48x4.U8.mu1.75.pattern.delta0.25.yedge1.out_eachm.pairing}}}
	\caption{
    	SC pairing order parameter on the vertical bonds at a fixed $y$, versus position $x$, for %for the filled stripes with 
	 a $48\times 4$ cylinder at $1/8$ hole doping, computed from DMRG with $U(1)$ symmetry ($S^z_\mathrm{tot}$) and two-site update.
		Pairing fields with 
		$h_p=0.25$ 
		are applied to the vertical bonds at the left edge $x=0$.
        Results with different bond dimension $m$ are shown.
        Both exponential (red line) and algebraic (blue curve) fits are shown;
        the solid (dashed) region indicates the region (not) used in the fits.
        The oscillation of the pairing order parameter coincides with the stripe period.
        The inset shows the (negative) pairing order parameters on the vertical ($\parallel$) (horizontal ($\perp$)) bonds for $m=12000$.
       % The DMRG truncation error for $m=12000$ is $\approx 10^{-5}$.
	}
	\label{fig:pair_order}
\end{figure}

\begin{figure}[h!]
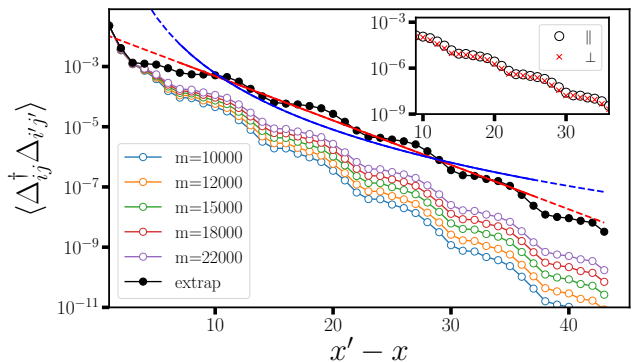
	
	\includegraphics[width=0.99\columnwidth]{{{U8_48x6_pcorr_extrap_loglog}}}
	\caption{
		%\C{({\em Added correlation length}.)} 
		SC pair-pair correlation on the vertical bonds %for the filled stripes 
	 at a fixed $y$, versus pair separation,
	for	a $48\times 6$ cylinder at $1/8$ hole doping, computed from DMRG with $U(1)\times SU(2)$ symmetry and single-site update.
        Results of different bond dimensions $m$ as well as the extrapolation to the infinite bond dimension are shown.
		The pair separation $(x'-x)$ is measured with respect to the reference bond, a vertical bond at $x=5$. 
        Both the exponential (red line) and the algebraic (blue curve) fits are shown;
        the solid (dash) region indicates the region (not) used in the fits.
        The correlation length is $\approx 2.9$ from the fitting.
        The inset shows the (negative) correlations on the vertical ($\parallel$) (horizontal ($\perp$)) bonds for $m=22000$.
    }
	\label{fig:pair_corr}
\end{figure} 

\begin{figure}[h!]
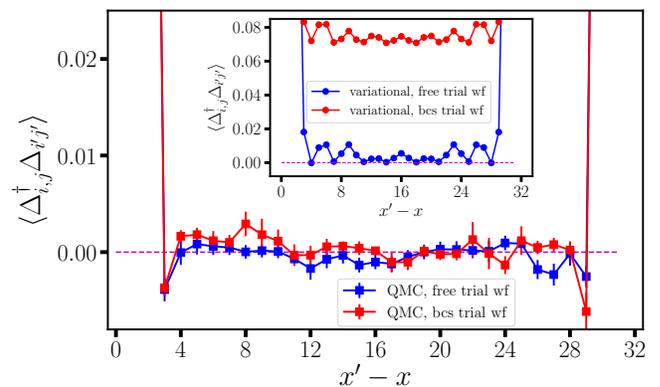
	
	\includegraphics[width=1.0\columnwidth]{{{BCS_vs_SLATER_U8_1o8doping_dwave_pairing}}}
	\caption{Pair-pair correlation from QMC for $32 \times 6$ system with $U = 8$, $1/8$ doping and PBC. 
		Different trial wave-functions are used. In the inset, the results from trial wave-functions
		themself are shown. Comparing to the variational result in the BCS wave-function, the
		pair-pair correlation in the QMC calculation using it as trial wave-function is largely
		suppressed.  
	}
	\label{QMC_8_32_PBC_pair_corr}
\end{figure}

We begin this section by considering a  $48\times 4$ cylinder at  $h = 1/8$, 
with a pairing pinning field  
applied on vertical bonds at the left edge only. We measure
how the SC pairing order parameter $\langle\hat{\Delta}_{ij}\rangle$ decays as a function of distance from the left edge, which gives an indication of the 
behavior of the pairing correlation function in the bulk.
We expect at least algebraic decay of the pairing order parameter if the system exhibits long range SC order, and exponential decay if there is 
no such order.
The calculations are done with DMRG, 
without conserving particle number  but with $U(1)$ symmetry for $S^z_\mathrm{tot}$.
In Fig.~\ref{fig:pair_order} we show the SC pairing order parameter on the vertical ($\hat y$) bonds along the
$\hat x$-direction.
The SC pairing order parameter is well converged when the bond dimension reaches $m=12000$, so no extrapolation is needed.
We perform both exponential and algebraic fits.
The SC pairing order parameter clearly decays exponentially versus distance from the pinning field. 
As shown in the inset in Fig.~\ref{fig:pair_order}, the pairing order on the horizontal ($\hat x$) bonds are perfectly symmetric with the negative values of the vertical bonds.
Although the pairing pinning fields are applied only on the vertical bonds at the left edge, the whole system
spontaneously builds a $d$-wave pairing structure.
This further confirms the tendency for short-range $d$-wave pairing, 
however the exponential decay of the pairing order indicates that there is no long-range SC order in the filled stripes.

We next investigate the pair-pair correlation directly on a $48\times 6$ cylinder.
At $h = 1/8$ doping, the ground state has  filled stripes with $\lambda=8$ \cite{Zheng1155}.
Previous study on width-2 ladders found that the SC pair-pair correlation decays algebraically\cite{PhysRevB.92.195139},
while further study on width-4 cylinder showed that the correlation decays exponentially \cite{PhysRevB.95.125125}.
Here we study a width-6 cylinder, 
employing the $U(1)\times SU(2)$ symmetry adapted DMRG with single-site update, and keeping bond dimension up to $22000$ $SU(2)$, 
to our knowledge the largest bond dimension to date in studying the SC pairing on width-6 cylinders. 
The computed SC pair-pair correlation is then extrapolated with
respect to the two-site energy variance \cite{PhysRevB.97.045125}.
(The extrapolation details can be found in the Appendix.)
Again we perform both exponential and the algebraic fits.
As shown in Fig.\ref{fig:pair_corr},
the SC pair-pair correlations follow clearly an exponential decay with pair separation, 
showing no
long-range order.
The inset shows the (negative) values of the correlations on the vertical (horizontal) bonds.
The correlations on the horizontal bonds are perfectly symmetric with the vertical bonds but
with opposite sign, again confirming the $d$-wave structure.

On the width-6 cylinders, in addition to the filled-stripes with $\lambda=8$, 
the $2/3$-filled stripes with $\lambda \approx 5$
can also be stabilized, and has a slightly higher energy
($\sim 0.001t$) \cite{Zheng1155} than the ground state studied above. 
This state has wavelength closer to the stripes
($\lambda=4$) observed experimentally \cite{nature_375_15_1995}, so 
it is interesting to also investigate the pairing in this meta-stable state.

We computed with DMRG the pair-pair correlation function in $48\times 6$ cylinders at $1/8$ hope doping in this state.
As in the ground state with filled stripes, the 
results are shown in Appendix~\ref{Sec:3-2_stripe}. 
The pair-pair correlation is found to decay exponentially, even faster than in the ground state with filled stripes.

It is worth emphasizing 
that we have used two different DMRG schemes above, 
one under $U(1)$ symmetry ($S^z_\mathrm{tot}$) with two-site updates, and the other under $SU(2)\times U(1)$ (spin and particle number) with strictly single-site updates. 
The consistency between the two approaches is a further confirmation of their reliability.
The width of the systems which can be studied efficiently with DMRG is limited due to the linear increase of entanglement entropy with
the width. To reach the TDL in two dimensions, we complement DMRG with two kinds of AFQMC calculations, computing both the pair-pair correlation function
and the pairing order parameter, as described next.

\begin{figure*}[t]
	\includegraphics[width=58mm]{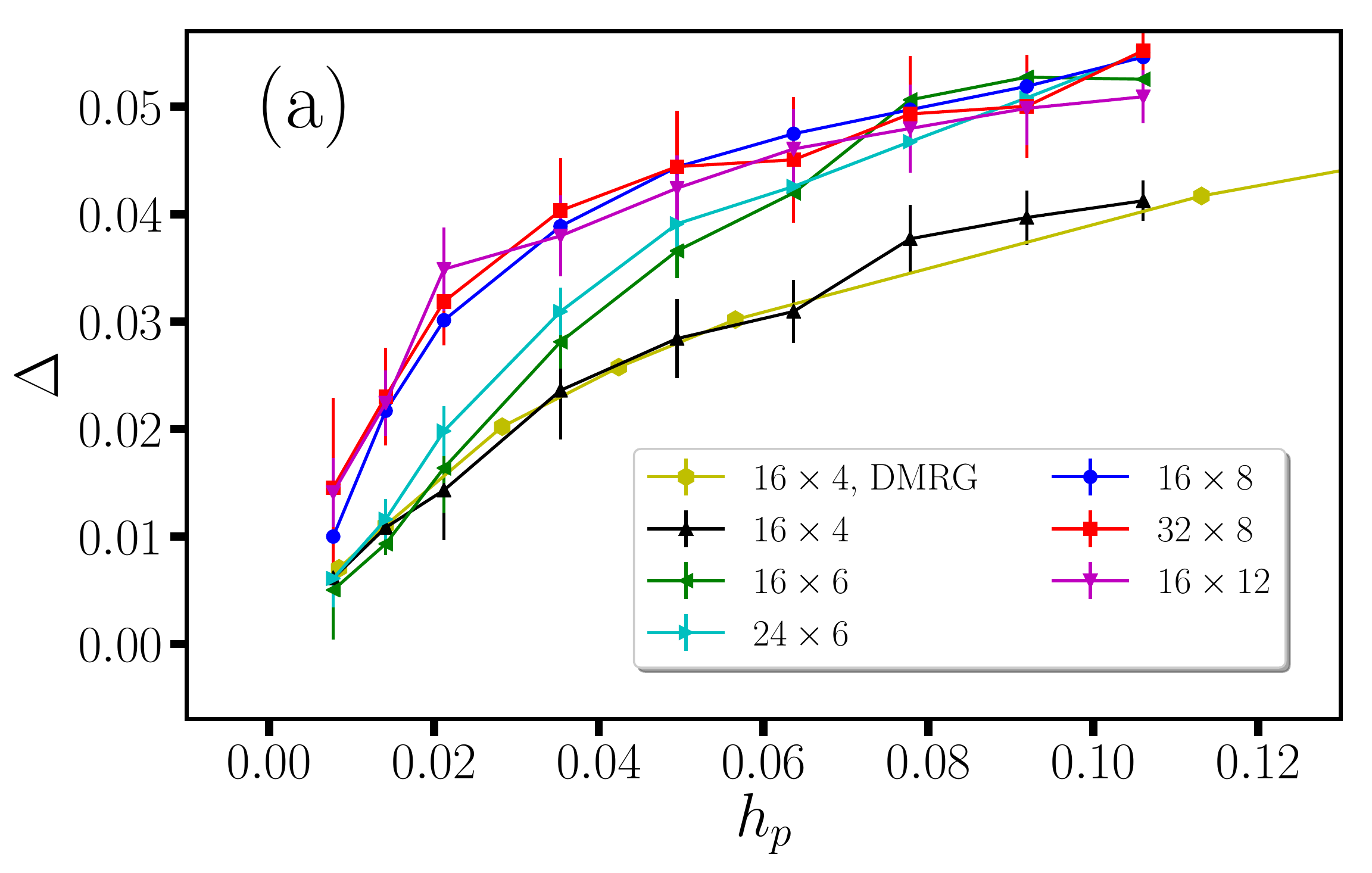}
	\includegraphics[width=58mm]{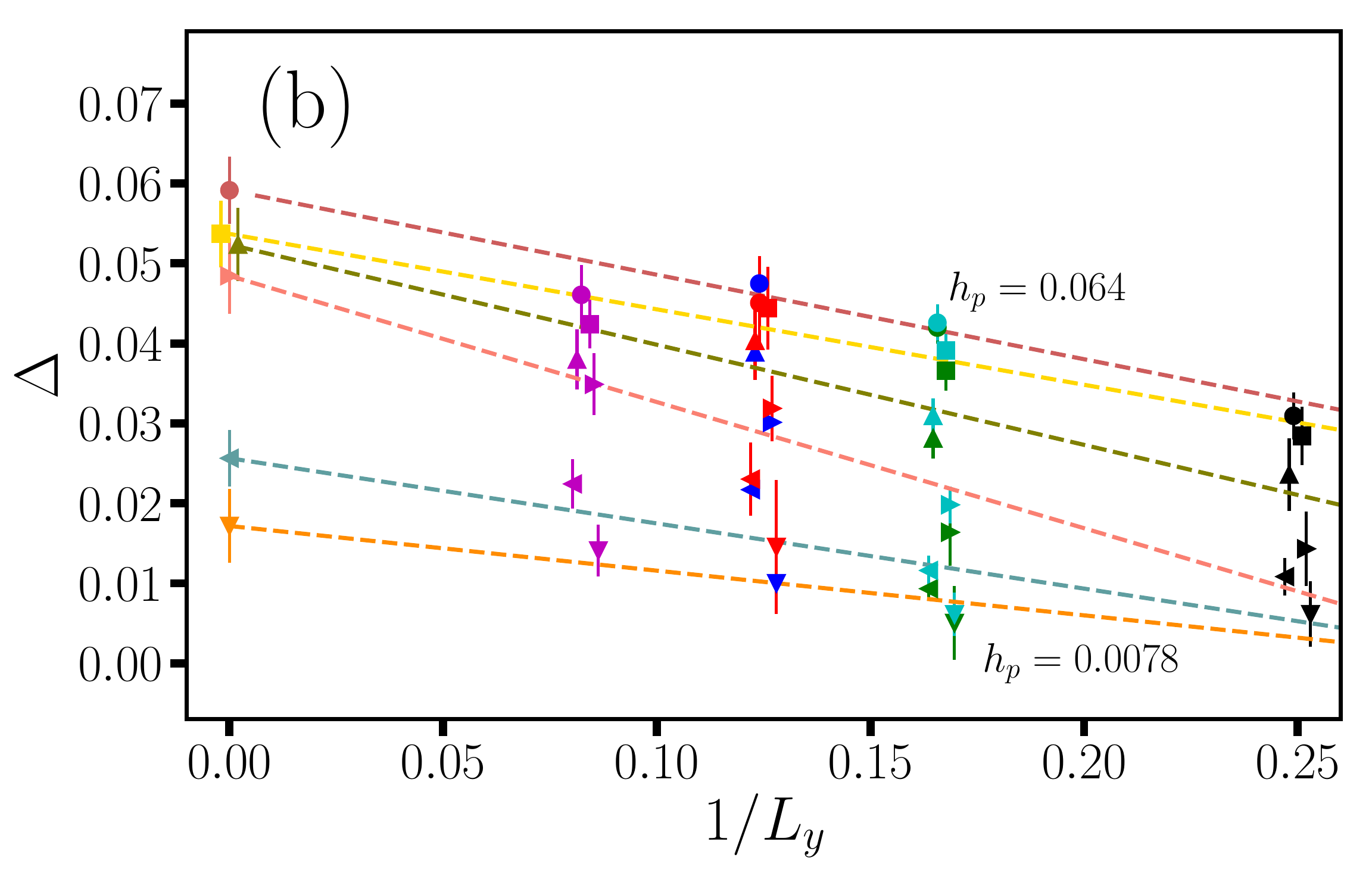}
	\includegraphics[width=58mm]{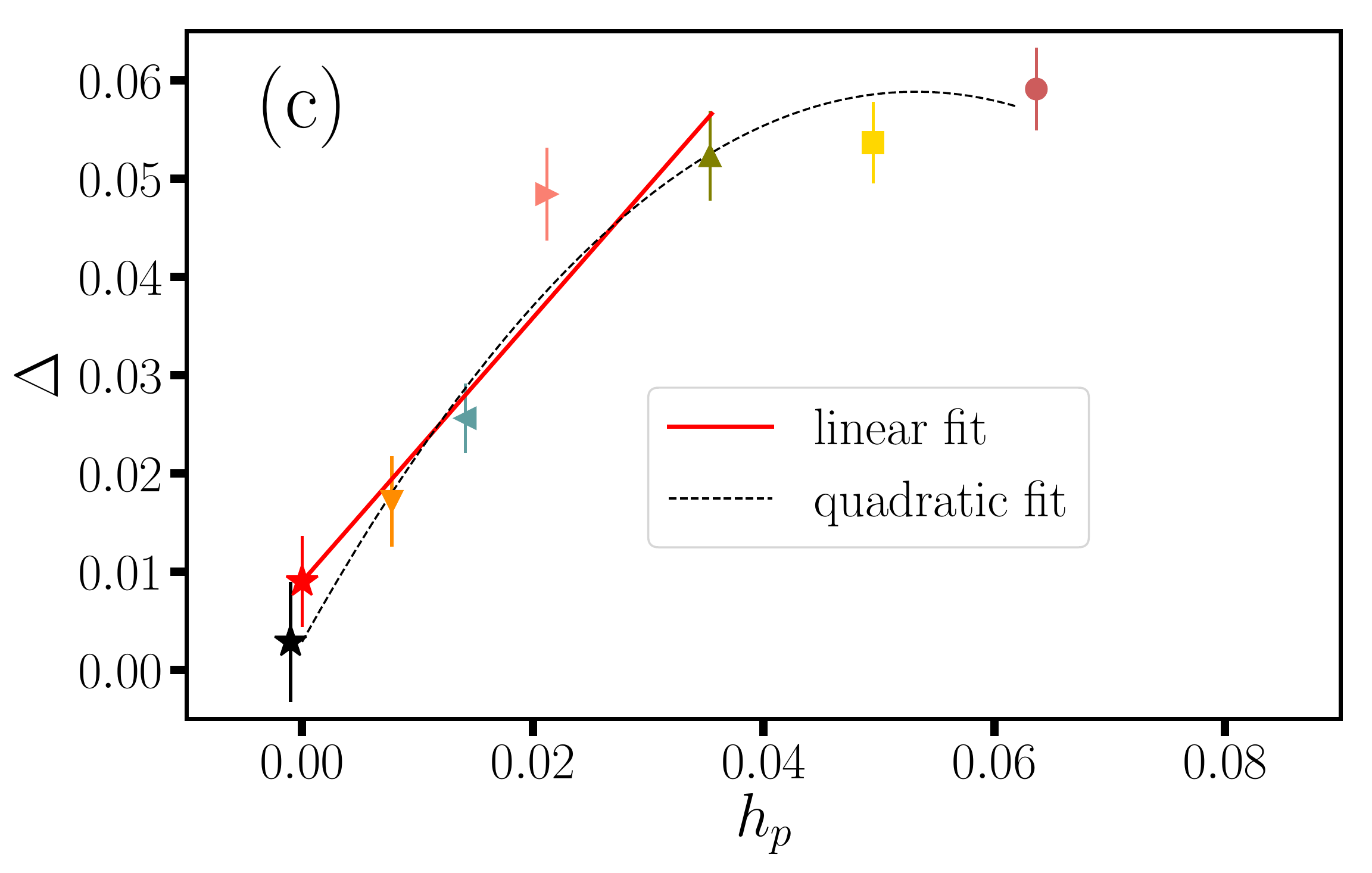}
	\caption{
		%\C{({\em We performed additional calculations to cut the statistical errors of the $24 \times 6$ results by half}.)}
		Finite size scaling of the SC pairing order parameter, at $U=8$ and $h=1/8$.
	 In (a), the computed pairing order parameters, $\Delta$, are shown for a variety of system sizes in the small $h_p$ region.
		In (b) an extrapolation of the results in (a) is performed with respect to the width in the periodic direction (linear in $1/L_y$) for 
		each value of $h_p$. 
		The lines, from top to bottom, are for $h_p=0.064, 0.049, 0.035, 0.021, 0.014$, and $0.0078$, respectively. 
		The resulting pairing order parameter $\Delta_\infty(h_p)$ is shown on the left. (A slight shift in the horizontal position 
		has been applied to some of the data points for better visibility of the results and error bars.) 
		In (c),  the result from the fit in (b), $\Delta_\infty(h_p)$, is plotted versus $h_p$.
		A quadratic fit is then performed, which yields a value $\Delta_\infty(0)= 0.003(6)$, as
		indicated by the black star at $h_p=0$. 
		A linear fit of the last few points is also shown.
		Weighted least square fits are used to account for the statistical errors.
	}
	\label{pair_size_scaling}
\end{figure*}

The pair-pair correlation function in an $32 \times 6$ supercell with PBC along both directions is shown
in Fig.~\ref{QMC_8_32_PBC_pair_corr}. This calculation is performed with fixed number of electrons, at $h=1/8$, with $U = 8$. 
The calculation with a free-electron trial wave function is consistent with earlier results from CP-AFQMC on square lattices \cite{PhysRevLett.78.4486,PhysRevB.59.1706}. 
The calculation with a number-projected BCS trial wave function, as mentioned, employed a new method \cite{PhysRevA.100.023621} which allows direct computation and 
back-propagation in the Hubbard model working in canonical ensemble.
For both free electron and BCS trial wave-functions, the pair-pair
correlations  from AFQMC are seen to decay to $0$ within the statistical resolution beyond a few lattice spacings.
The BCS wave-function itself has very large pair-pair correlation, as shown 
	in the inset. %of Fig.~\ref{QMC_8_32_PBC_pair_corr}. 
	However in the AFQMC
result using it as trial wave-function, the pair-pair correlation %(the red line in the main plot of Fig.~\ref{QMC_8_32_PBC_pair_corr})
is %largely 
suppressed by two orders of magnitude, and decays to $0$ beyond a few lattice spacings. While the agreement between the two trial 
wave functions is not perfect, their consistent behavior at large pair separation provides another corroboration of the results from the pairing order parameter.

We next employ AFQMC to
calculate the pairing order parameter, by applying $\hat{H}_p$ as in Eq.~(\ref{eqn:Hp}),
with the 
pairing fields chosen to  match the $d_{x^2-y^2}$ structure and applied
throughout the supercell.
The pairing order parameter
$\Delta$ (averaged over all bonds) 
is calculated as a function of $h_p$. 
The chemical potential $\mu=1.75$ is chosen such that the hole density is $h = 1/8$  in the ground state when $h_p=0$, and held fixed for 
all $h_p$.
To detect possible long-range SC pairing order in the pure Hubbard model ($h_p=0$), we need to reach the TDL at each $h_p$
first, then extrapolate $h_p$ to zero. This procedure is parallel to what is illustrated in Fig.~\ref{mag_size_scaling}, and is 
 shown in
 Fig.~\ref{pair_size_scaling}.
Following the procedure discussed in Sec.~\ref{subsec:2measures},
AFMQC calculations are performed on various system sizes up to $32\times 8$.
We focus on the small $h_p$ region where the behavior determines whether long-range SC pairing
order exists. 
In this region the self-consistent trial wave function gives the same results as 
the non-interacting trial wave-function; the latter form is used here, obtained by setting $U = 0$
in Eq.~\ref{eqn:H_pairing} and tuning the chemical potential to give a doping of $h = 1/8$.
%\C{The non-interacting trial wave-function gives smaller fluctuation (statistic error), which saves
%computational time especially for large system size.}
The computed pairing order parameters as a function of $h_p$ are shown in Fig.~\ref{pair_size_scaling} (a).
For the lengths studied here ($L_x >16$),
the results are 
not sensitive to $L_x$,
as can be seen by comparing the $16\times 6$ and $24\times 6$ results, and also the $16\times 8$ and $32\times 8$ results.
On the other hand, the order increases when the system becomes wider,
although the dependence on system size becomes weak beyond $L_y=6$.
To obtain the order parameters at the TDL, $\Delta_\infty(h_p)$, a linear extrapolation with $1/L_y$ %to the width 
is performed for each value of $h_p$, as shown in Fig.~\ref{pair_size_scaling}(b).
The resulting $\Delta_\infty(h_p)$ and the statistical uncertainties, are shown in Fig.~\ref{pair_size_scaling}(c). 
A quadratic fit is then performed, which yields a value $\Delta_\infty(0)= 0.003(6)$, as
indicated by the symbol at $h_p=0$.
A linear fit for $h_p < 0.05$ is also shown, which gives a statistically consistent result. 
We thus conclude that there is no long-range SC pairing in this system in the TDL.

\begin{figure}[t]
	\includegraphics[width=80mm]{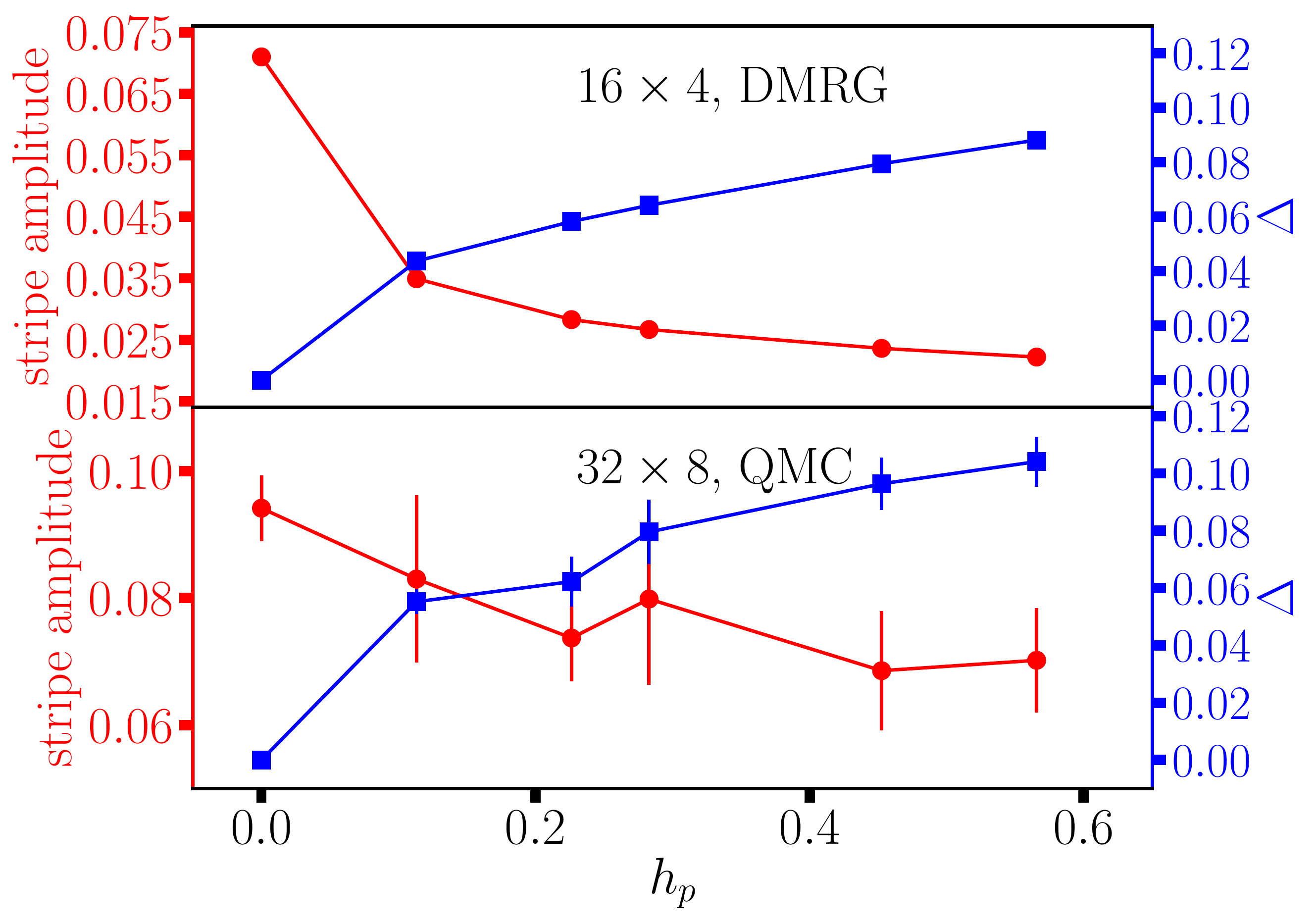}
	\caption{
	%\C{({\em Added data at $h_p=0$}.)}
		Strengths of the stripe and paring orders versus the applied pairing field, $h_p$. 
	The upper and lower panels show results from  $16 \times 4$  and  $32 \times 8$ cylinders, respectively, both at $U=8$ with $h=1/8$.
	}
	\label{stripe_pair}
\end{figure}

\begin{figure}[t]	
	\includegraphics[width=80mm]{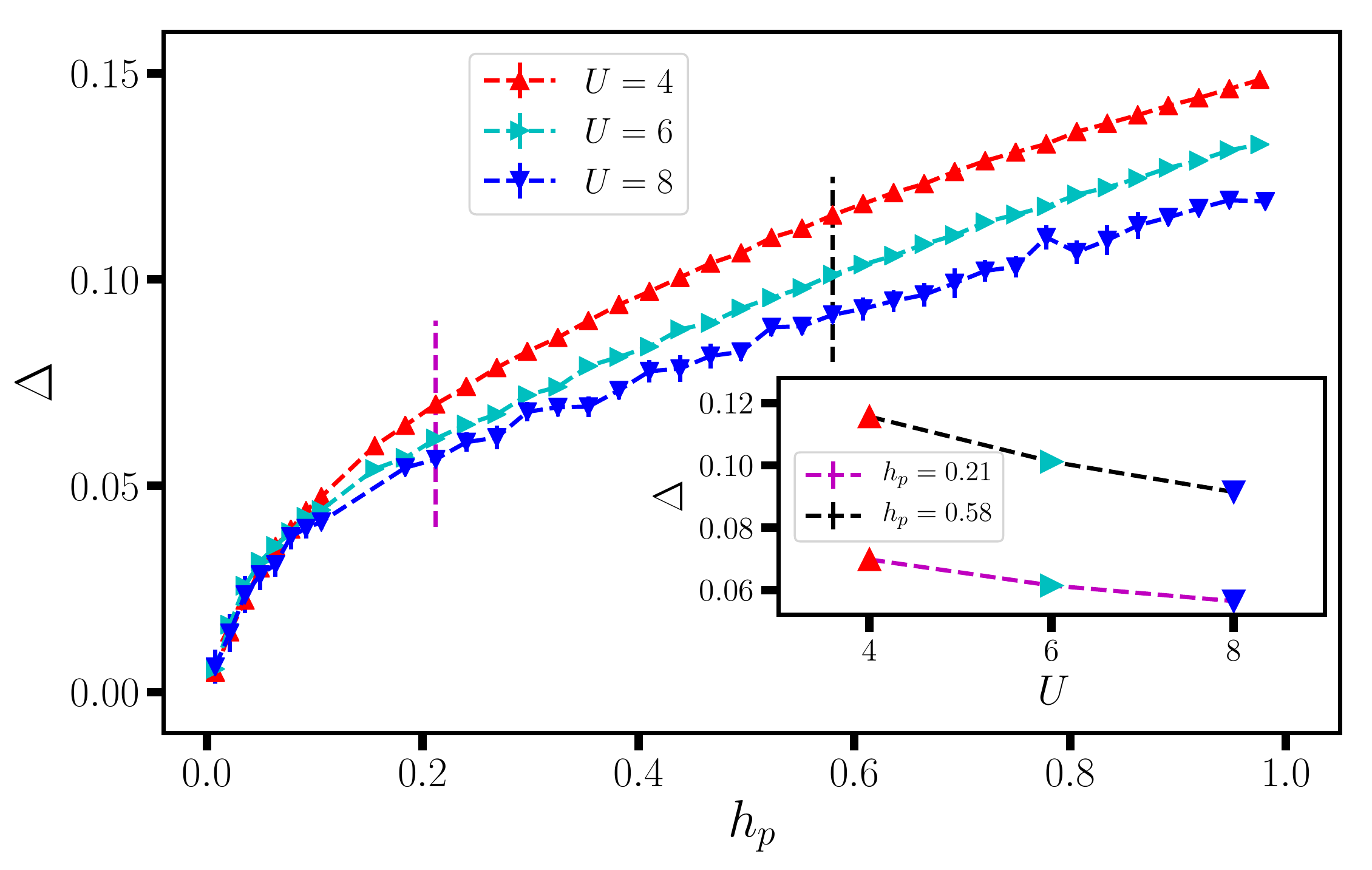}
	\caption{Dependence of the pairing order parameter on interaction strength $U$. 
	Calculations were done in the same manner as in Fig.~\ref{E_pair_4_16}, on $16\times 4$ systems, varying only $U$. 
	In the inset the  pairing order parameters are plotted versus $U$ for two values of $h_p$ as indicated by the vertical lines in the main graph, $h_p = 0.38$ and $0.58$. 
		 }
	\label{pair_U}
\end{figure}

\begin{figure}[t]	
	\includegraphics[width=80mm]{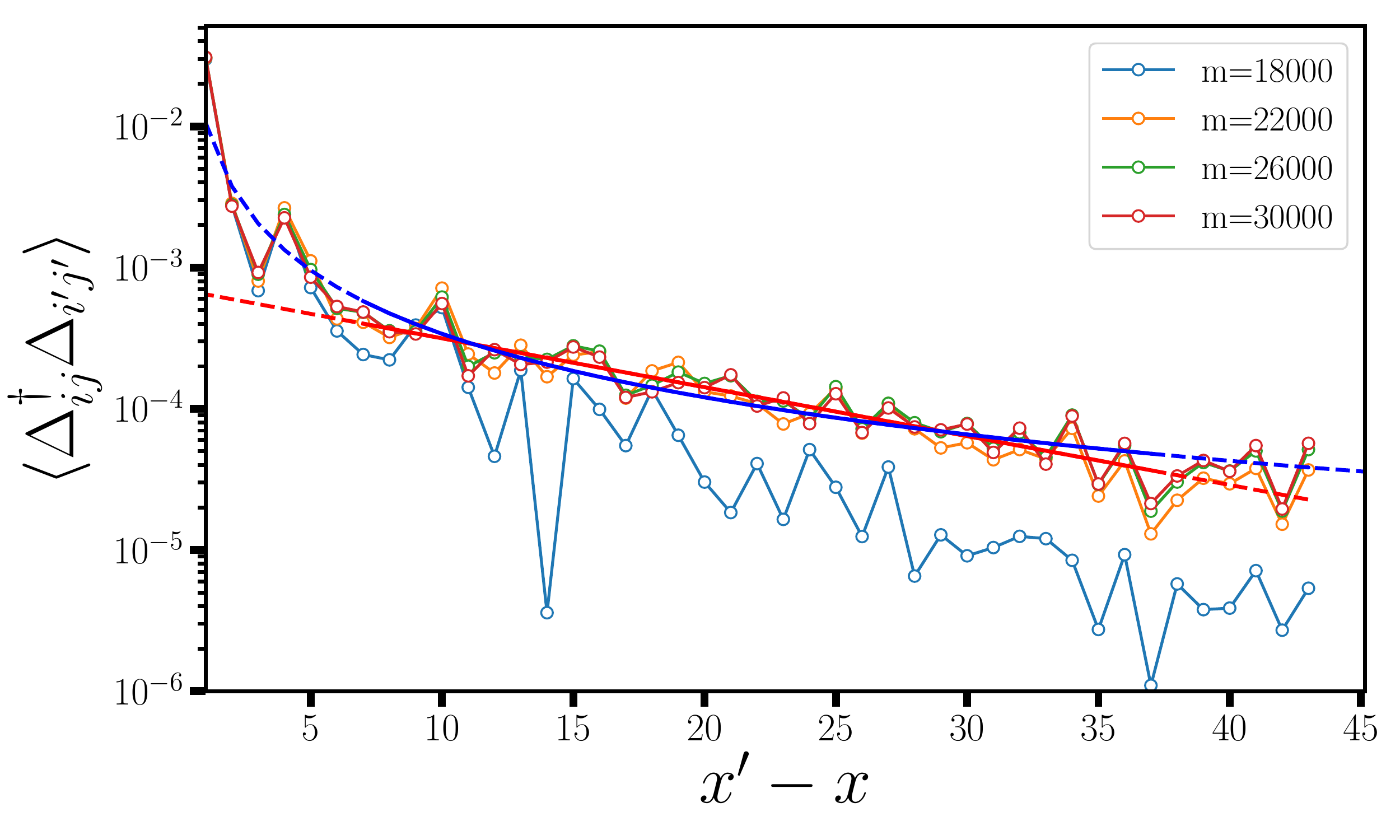}
	\caption{SC pair-pair correlation on the vertical bonds for $U=4$ at $h=1/8$ on a $48\times 4$ cylinder, computed by DMRG with $U(1)\times SU(2)$ symmetry and the  single-site update.
        The reference bond is a vertical bond at $x=5$.
        Both exponential (red line) and algebraic (blue curve) fits are shown;
        the solid (dash) region indicates the region (not) used in the fits.
		}
	\label{pair_corr_U4}
\end{figure}

\begin{figure*}[t]	
	\includegraphics[width=58mm]{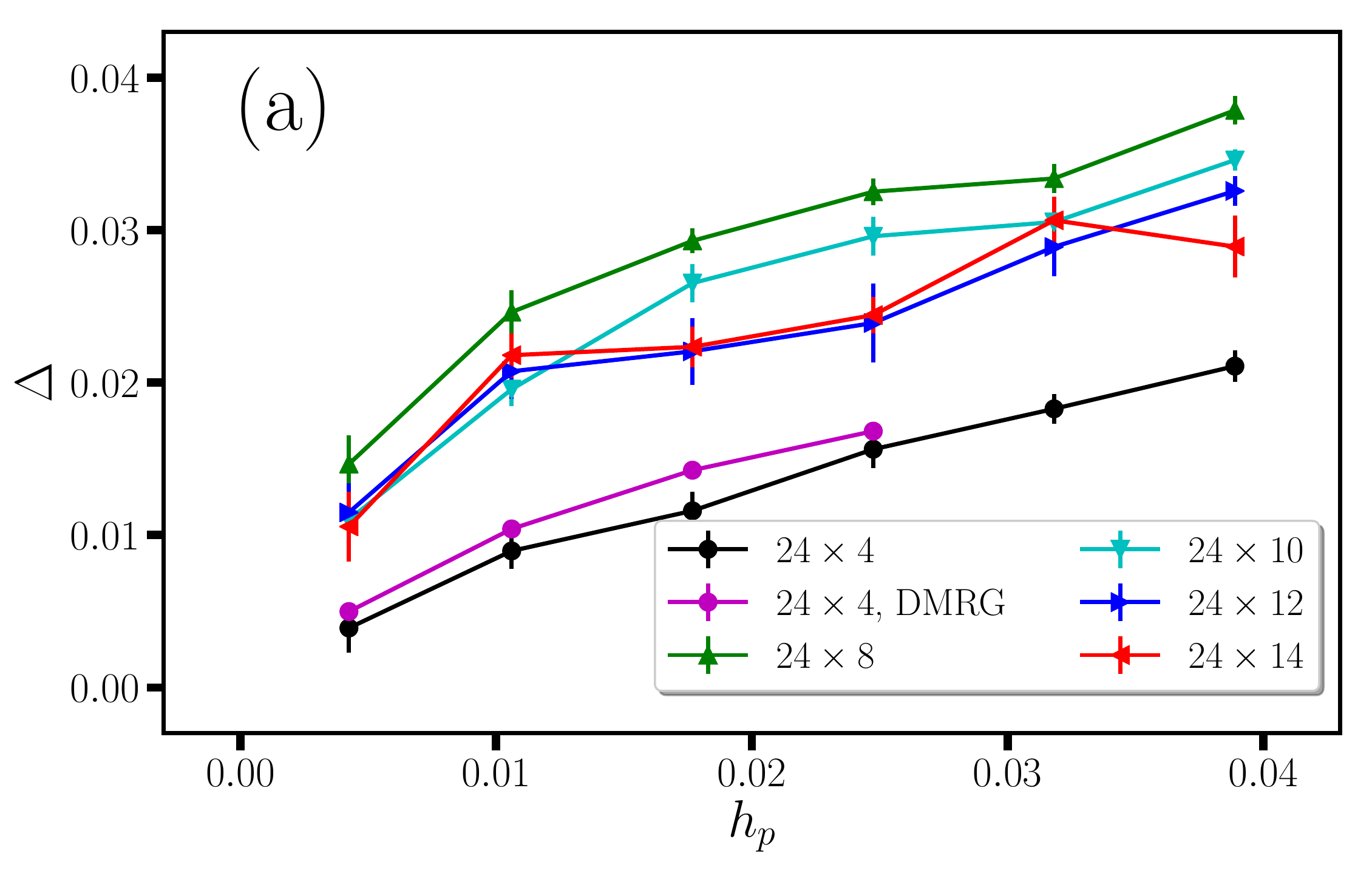}
	\includegraphics[width=58mm]{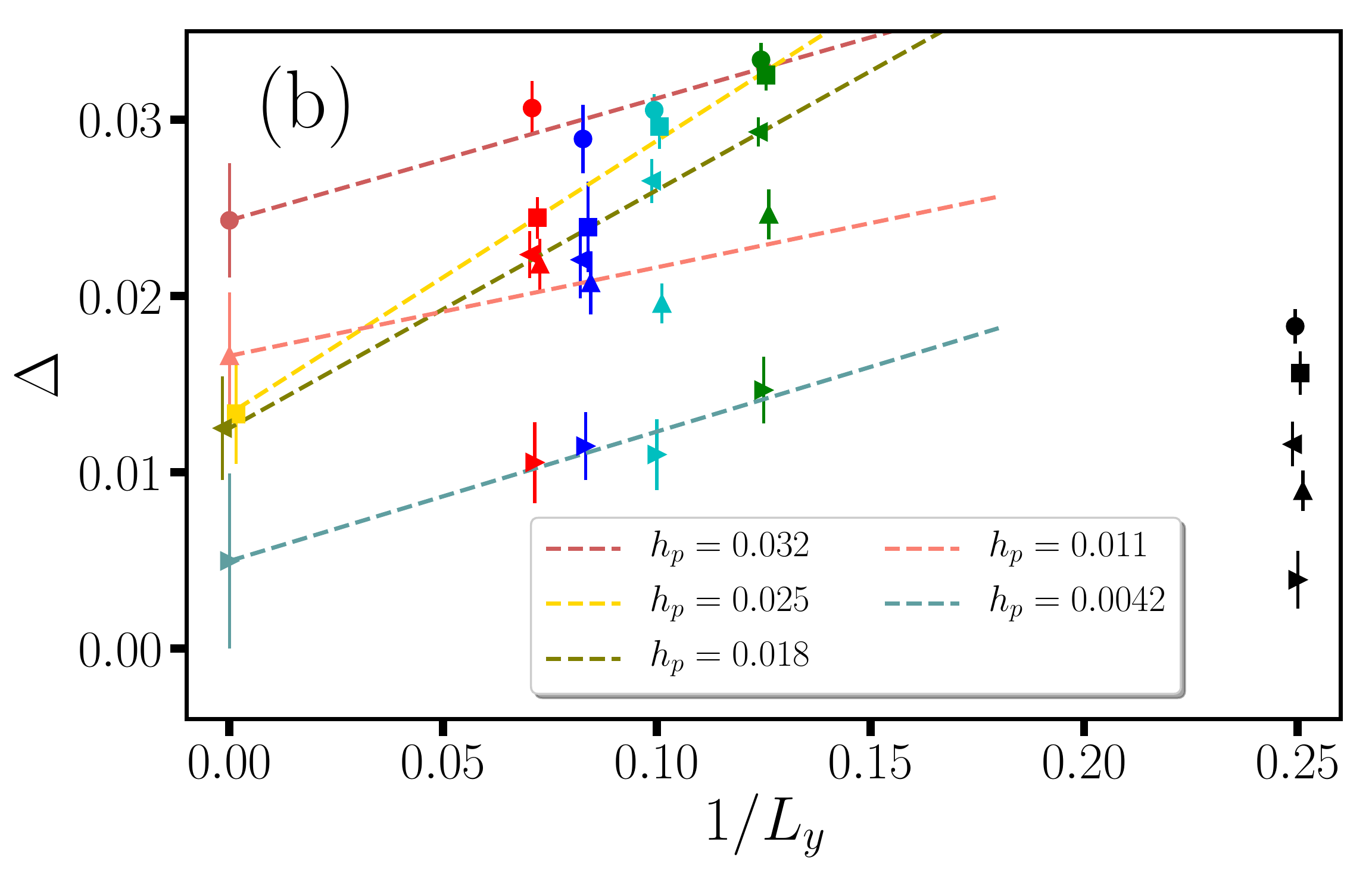}
	\includegraphics[width=58mm]{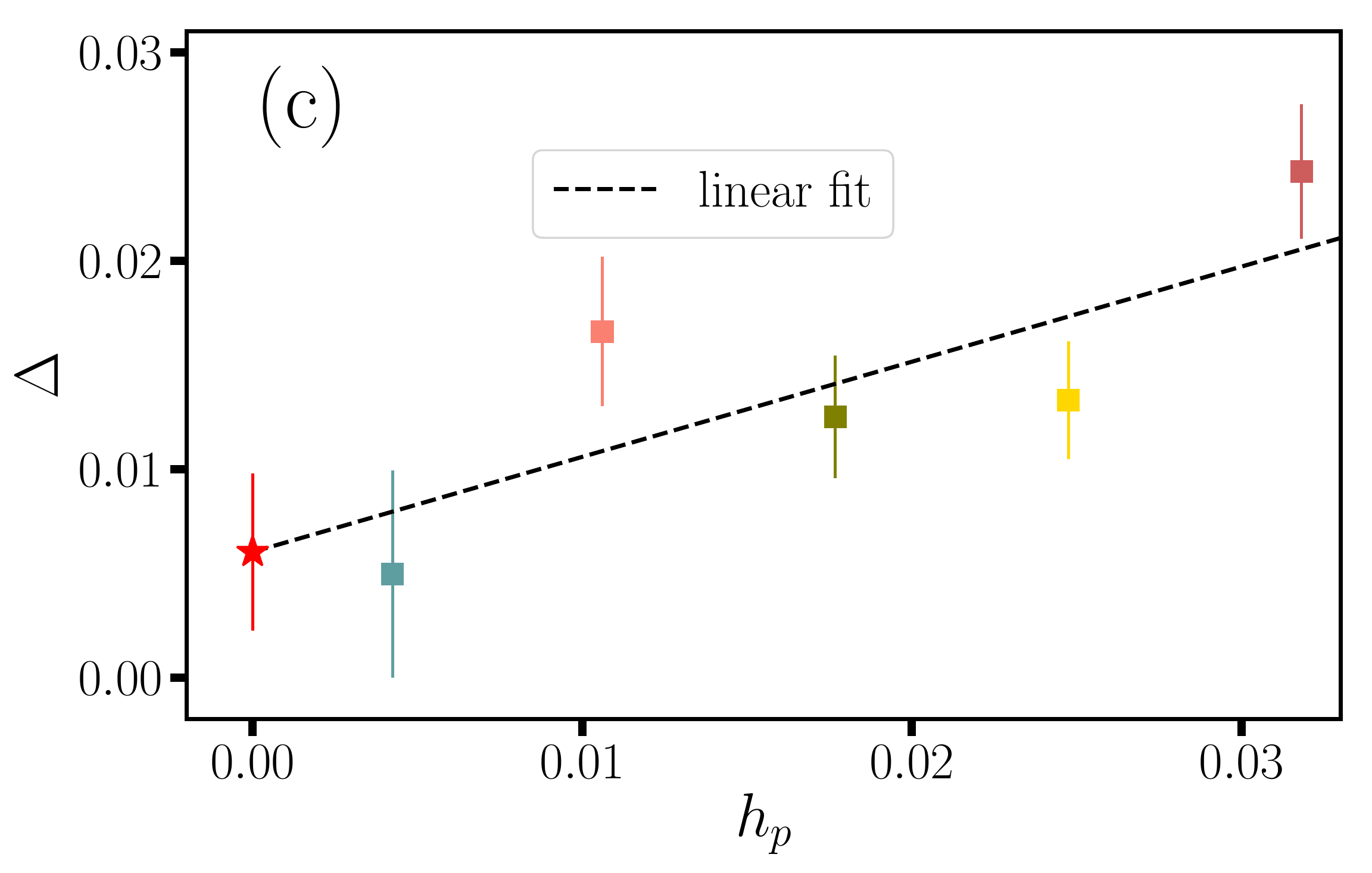}	
	\caption{Finite size scaling of the SC pairing order parameter, at $U=4$ and doping of $h \simeq 1/6$. 
	The layout is the 
	same as in Fig.~\ref{pair_size_scaling}. 
	In (a) the pairing order parameter computed from DMRG  for $24 \times 4$ (see Fig.~\ref{4_24_1_6_U4_benchmark}) 
		are also shown for comparison.
		  In (b), the width-4 data (black symbols) were not included in the linear fits, as they are non-monotonic with larger $L_y$ results.
		In (c), the extrapolation 
		yields a value $\Delta_\infty(0)= 0.006(4)$, as indicated by the symbol at $h_p=0$.
		}
	\label{pair_U4_1_6}
\end{figure*}

\subsection{The competition between stripe and superconducting orders}
\label{ssec:stripevSC}

In Fig.~\ref{stripe_pair}, we examine the trend of the stripe and SC pairing orders when the strength of the applied pairing field is varied.
%\C{The existence of filled stripe order in the ground state was established in Ref.~\cite{Zheng1155}}
The stripe order amplitude is defined as
 the intensity of hole modulation in the stripe state (i.e., the maximum value minus the minimum value of hole density along the longer direction of cylinder). 
Results are presented for
two systems, a $16 \times 4$ cylinder computed by DMRG and a $32 \times 8$ cylinder by AFQMC.  Both results are for $U = 8$, with
$h = 1/8$. 
We can see that, 
when  $h_p$ is decreased,  
		the pairing order parameter becomes smaller, while the stripe order increases. 
The two orders thus compete against each other in the Hubbard model. 
The results of the previous 
subsection 
show that, at the zero pairing field limit, the stripe order dominates and no pairing order survives  in this parameter region.

\subsection{$U$ dependence, and the case of $U = 4$} 
\label{ssec:U4}

In this section we study the pairing order at different interaction strengths $U$. 
In Fig.~\ref{pair_U}, we plot the pairing order parameter for $U = 4, 6$ and $8$ at $1/8$ doping. 
These calculations follow the same procedure as in Fig.~\ref{E_pair_4_16}. performed on  
a $16 \times 4$ system, with
 pairing fields of $d$-wave structure applied to the entire system. In
the large $h_p$ region,
we find that the SC pairing order parameter increases as $U$ decreases. 
In the small $h_p$ region, the pairing
order parameter varies little 
with the decrease of $U$.

Given the tendency for the pairing susceptibility to increase as $U$ is reduced, we next focus on a lower but still physically relevant value of $U=4$.
In Fig.~\ref{pair_corr_U4}, we show the pair-pair correlation function for $1/8$ doping from DMRG.
This study at  $U = 4$ is similar to the one in Fig.~\ref{fig:pair_corr} at $U=8$. 
Consistent with the result from Sec.~\ref{ssec:stripevSC}, the pairing correlation amplitude is substantially larger than for $U=8$.
Both exponential and algebraic fits are performed to the correlation function computed with the largest bond dimension kept, $m=30,000$.
Here the results are less definitive.
The exponential is a slightly better fit to the data, but the algebraic fit cannot be ruled out conclusively. 

We next study the pairing order parameter at $U = 4$, using AFQMC to approach the TDL. We target $h = 1/6$, near optimal doping.
In this parameter regime, 
no stripe or spin-density wave state is observed in the ground state of the Hubbard model \cite{PhysRevLett.104.116402}. 
This is the system where one of the extensive cross-checks between DMRG and AFQMC was performed in Sec.~\ref{subsec:2measures}
(Fig.~\ref{4_24_1_6_U4_benchmark}).
The computed pairing order parameter for a variety of system sizes are shown in Fig.~\ref{pair_U4_1_6}. 
The same
procedure as in Fig.~\ref{pair_size_scaling} is performed. We first carry out an extrapolation with the width $L_y$ at
each $h_p$, to reach the TDL.
In contrast to the $U = 8$ case, the pairing order parameter here is seen to either decrease or saturate very quickly with system width  
as $L_y$ grows. 
The results for width-4 cylinders ($24 \times 4$) are seen to be non-monotonic with wider systems, so they  are not included in the fit. % with $1/L_y$.
The final result
extrapolated to the $h_p = 0$ limit is $\Delta_\infty(0)= 0.006(4)$, statistically compatible with a vanishing 
order parameter.

Of course, based on this and the pairing correlation results above, 
we can not fully rule out the possibility of  a finite pairing order in the ground state at 
$U=4$. (See also discussion on weak coupling in Sec.~\ref{sec:intro}.)
Our results do put a rather stringent bound on the strength of the pairing order,
which is considerably smaller than indicated by the best previous calculations with affirmative results on superconductivity.
The small
magnitude of this bound
suggests that, even if the pure Hubbard model 
is superconducting in some regime of the parameter space further from the most relevant physical parameterrs, it is likely missing key 
ingredients as a fundamental model for cuprate superconductors.

\section{summary and perspective}
\label{sec:summary}

In summary, we have carried out a %careful and 
detailed study of the superconducting pairing properties in the ground state of the 2D pure Hubbard model,
using two of the most accurate ground-state many-body computational methods available at present. With both methods, we have 
presented technical advances which enabled new capabilities in probing the superconducting order. 
The DMRG calculations of pairing correlation functions 
were performed on up to width-6 cylinders, with unprecedented accuracy. 
The AFQMC computations were, for the first time, able to compute 
pairing order parameters relying on total energy calculations.
Meticulous comparisons were made between the two methods. 
Their complementary application allowed us to maintain high accuracy and reach the thermodynamic limit.

In the parameter regime relevant to the cuprates ($U\sim 6$-$8$) we find that
the pure Hubbard model does not have a superconducting ground state.  We also
find that the lack of superconductivity is due to a competition with stripe
order, with stripes dominating.
At smaller $U \sim 4$, the tendency for striped ground states is much weaker.
In this case, we still find a pairing response consistent with zero. While we cannot rule out a small nonzero
pairing order parameter, %although %we can put bounds on it.
our results place an upper bound to its strength which is very small.

In the early days of high temperature cuprate superconductivity, when no numerical approach was adequate to accurately probe the low
temperature, doped regime in  large system sizes, it was natural to expect that if one could get past the fermion sign problem, one would
quickly have a clear picture of the physics involved. Now that we can study this regime, new obstacles have been revealed.  A
key obstacle is the close competition of a number of different phases, with small Hamiltonian terms mediating which phase
is favored. This situation makes simulations more difficult, but equally important is that it is very difficult to know reliably
which sets of small terms and parameters (such as next nearest neighbor hopping $t'$) describe the actual materials.
Our work can be viewed as a key initial step, where the iconic simplest-to-define model with only $U$ and $t$ is found not
to exhibit superconductivity. To go beyond this, one will want to study phases and superconductivity in generalized
models including a broad range of parameters. Simultaneously, it is important to improve our techniques for both deriving
accurate models, and for simulating real systems with very strong correlation without introducing models.

It is also important to note that 
the pure Hubbard model does get much of the physics right, including antiferromagnetism and its destruction upon doping, and a
tendency for stripes to occur and to compete with $d$-wave superconductivity. It also 
produces the crucial physics that there are many intertwined states separated by tiny energy scales, a key part of the reason 
that the complete nature of superconductivity in the cuprates remains to be resolved.

\begin{acknowledgments}
We acknowledge helpful discussions with E.~Gull, 
M.~Imada, and A.~J.~Millis. 
MQ and SZ were supported by the Simons Foundation. The Flatiron Institute is a division of the Simons Foundation.
SRW acknowledges the support of the NSF through DMR-1812558.
C.-M. C. and U.S. acknowledge support by the Deutsche Forschungsgemeinschaft (DFG, German Research Foundation) under
Germany's Excellence Strategy 426 EXC-2111 390814868.
Part of the calculations were carried out at
the Extreme Science and Engineering Discovery Environment
(XSEDE), which is supported by National Science Foundation grant number ACI-1053575,
and the computational facilities at the College of William and Mary.
C.H. acknowledges funding through ERC Grant QUENOCOBA, ERC-2016-ADG
(Grant no. 742102).
\end{acknowledgments}

\bibliography{Pairing-Hubbard.bib}

\appendix

\section{Particle-hole transformation}
When the pairing fields in Eq.~(\ref{eqn:Hp}) are applied, the Hamiltonian 
contains fluctuations of the total
particle number. 
The usual ground-state AFQMC is formulated in canonical ensemble with fixed $N_e$.
However, we can apply a partial particle-hole transformation 
as follows:
\begin{eqnarray}
\hat{c}_{i\uparrow} & \rightarrow & \hat{d}_{i\uparrow}, \quad \hat{c}_{i\uparrow}^{\dagger}  \rightarrow  \hat{d}_{i\uparrow}^{\dagger}\\
\hat{c}_{i\downarrow} & \rightarrow & \hat{d}_{i\downarrow}^{\dagger}(-1)^{i}, \quad \hat{c}_{i\downarrow}^{\dagger} 
\rightarrow  \hat{d}_{i\downarrow}(-1)^{i}\nonumber\,. 
\label{eq:partial_ph_trans}
\end{eqnarray}
Then the Hamiltonian in Eq.~(\ref{eqn:H_pairing}) 
is transformed to
\begin{equation}
\hat{H} = -t\sum_{\langle i,j\rangle\sigma}\hat{d}_{i\sigma}^{\dagger}\hat{d}_{j\sigma} + U\sum_{i}(\hat{m}_{i\uparrow}-\hat{m}_{i\downarrow}\hat{m}_{i\uparrow})
-\mu\sum_{i}(\hat{m}_{i\uparrow} + 1 - \hat{m}_{i\downarrow})\,, 
\label{eqn:H_pairing_ph}
\end{equation}
where $\hat{m}_{i,\sigma}=\hat{d}_{i,\sigma}^{\dagger}\hat{d}_{i,\sigma}$. The pairing operator in Eq.~(\ref{eqn:Hp}) is transformed from Eq.~(\ref{eqn:Delta_def})
to: $\hat{\Delta}_{ij}=((-1)^{j+1}\hat{d}_{j\downarrow}^{\dagger}\hat{d}_{i\uparrow}-(-1)^{i}\hat{d}_{i\downarrow}^{\dagger}\hat{d}_{j\uparrow})/\sqrt{2}$,
which now describes  spin-flip hopping terms.
%Now the pairing field term is transformed to a spin-flip hopping term. 
The chemical potentials for electrons with up and down
spin are now $\mu - U$ and $-\mu$ respectively, introducing spin imbalance in the system. The sign of interaction strength $U$ is
flipped which means the interaction turns to attractive after the transformation. In CP-AFQMC calculation, walkers (Slater determinants) are
now represented as $2N \times N_e$ matrix \cite{PhysRevB.94.085103} and each orbital in the Slater determinant is now a spin-orbital with a mixture of up and down orbitals. This is similar to the treatment of Hamiltonians with spin-orbit coupling terms \cite{ROSENBERG2017}.
Other details remain unchanged in the CP-AFQMC calculation.

\section{Comparison of energies between pairing state and stripe state}

\begin{figure}[t]
	\includegraphics[width=80mm]{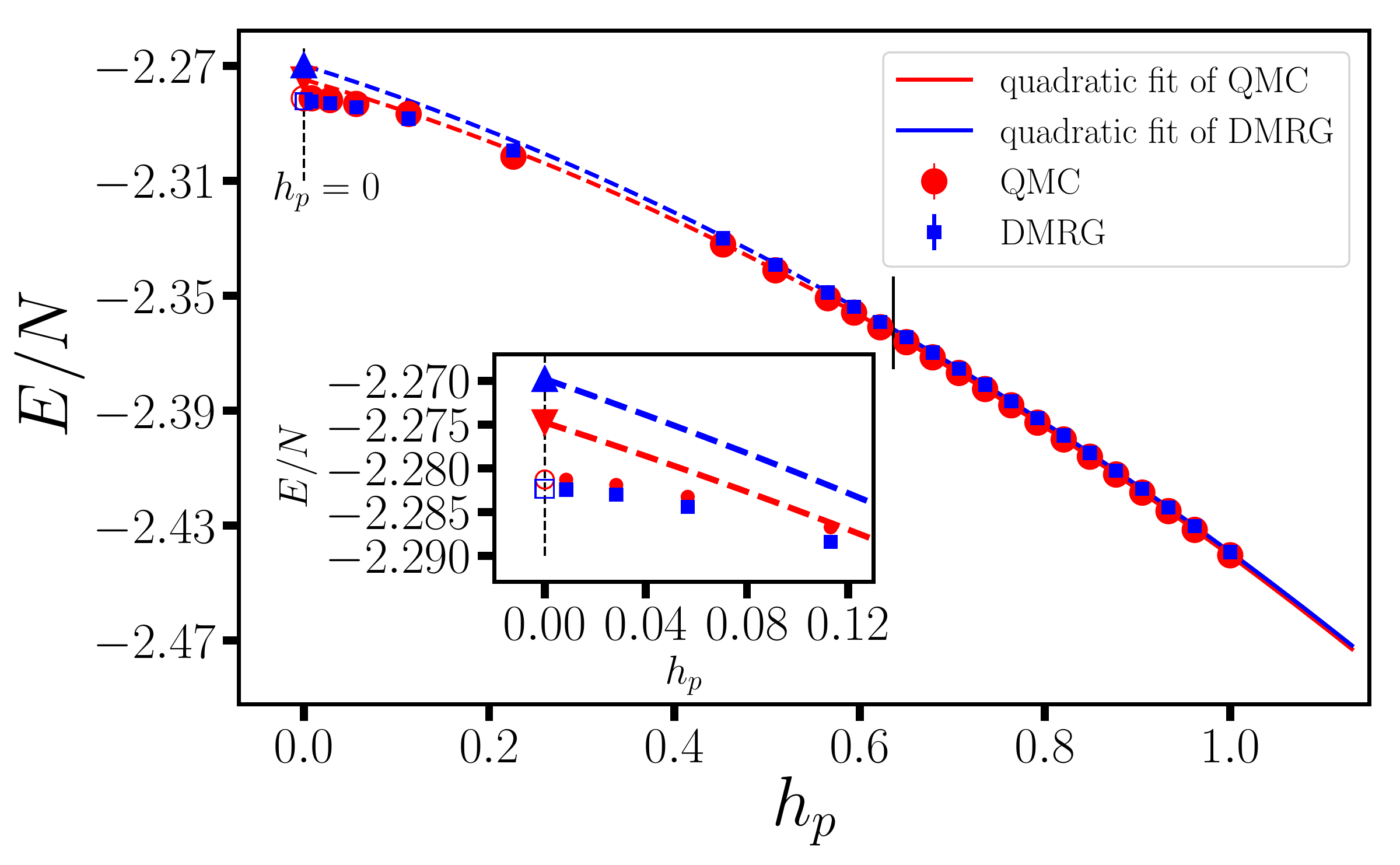}
	\caption{ 
		Comparison of the ground-state energies from CP-AFQMC (red) and DMRG (blue)
        , as a function of the pairing field strength $h_p$. The system is 
		a $16 \times 4$ cylinder. A fixed value of $\mu = 1.75$ is used with which the doping is $1/8$ when $h_p = 0$. 
		A quadratic fit is applied to each set of energies using only values at large $h_p$ (points to the 
		right of the vertical bar). The triangular symbols at $h_p=0$ show the intercept result from 
		the fit, while the open symbols are those obtained from actual calculations done at $h_p=0$.
		In the inset, a zoom of the main plot near $h_p=0$ is shown.
		).
		}
	\label{fig:E_pair_4_16_app}
\end{figure} 

In Fig.~\ref{fig:E_pair_4_16_app} we show the comparison of energies between pairing state and stripe state.
The systems are $16\times 4$ cylinders with $d$-wave pairing field applied on the whole systems.
The energy for pairing state at $h_p = 0$ (denoted by triangles in Fig.~\ref{fig:E_pair_4_16_app}) is obtained from a quadratic fit with energies at large $h_p$. The stripe energy (denoted by open square and cycle in
Fig.~\ref{fig:E_pair_4_16_app})) is the actual value calculated at $h_p = 0$. We can find the energy of pairing state
is slightly higher than that of stripe state, by about $\sim 0.01$ per site.

%-----------------

%\section{Doping dependence on 4-leg cylinders}
\section{Further investigation of doping dependence}
\label{Sec:supp_doping}
\begin{figure}[h!]
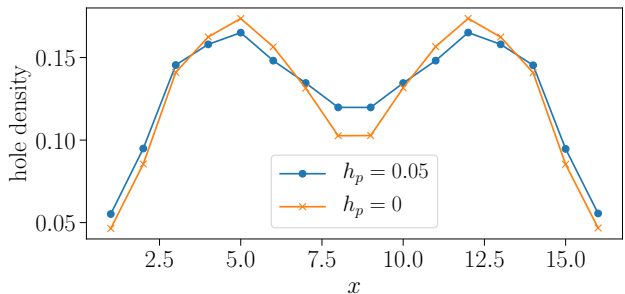
	
\includegraphics[width=0.99\columnwidth]{{{compare_hole}}}
\caption{
    Comparison of the local hole densities on different rungs for $16 \times 4$ systems with $h_p=0.05$ and with zero pairing field.
    The system with pairing field is the same as in Fig.~\ref{fig:pair_vs_n} with $\mu=1.72$.
    }
    \label{fig:compare_hole}
\end{figure}

We have shown in the main text the pairing order induced by the pairing field $h_p=0.05$ for different doping in Fig.~\ref{fig:pair_vs_n}.
Here we illustrate that the pairing field amplitude $h_p=0.05$ only slightly changes the original ground state and thus the induced SC order properly represents the response of the original ground state.
In Fig.~\ref{fig:compare_hole} we compare the local hole densities of two $16 \times 4$ systems.
One is a system with pairing field $h_p=0.05$ and $\mu=1.72$, corresponding to overall particle density $\approx 0.872$.
The other is a system of a conserved particle density $0.875$ without pairing field.
The stripe order can be clearly seen in both systems, and the pairing fields only slightly change the local densities.

\begin{figure}[h!]
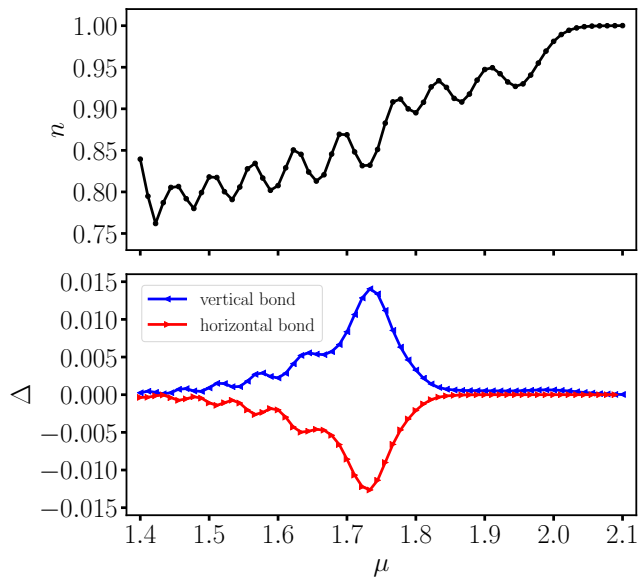
	
\includegraphics[width=0.99\columnwidth]{{{n_delta_mu}}}
\caption{
    The system is a $64\times 4$ cylinder with local chemical potential $\mu$ linearly changing along the longitudinal direction.
    (Upper panel) Particle densities along the longitudinal direction. The $x$-axis has been replaced by the local $\mu$ on the corresponding position.
    (Lower panel) The local SC orders on the vertical (blue) and horizontal (red) bonds along the longitudinal direction.
    The results are obtained by using DMRG without conserving total particle number (grand canonical ensemble).
    }
    \label{fig:doping_denp}
\end{figure}

To further examine the doping dependence of the pairing order, we study by DMRG a $64\times 4$ cylinder with local chemical potential $\mu(x)$ linearly changing from $1.4$ to $2.1$ along the longitudinal ($x$) direction.
Since the local density will vary with %in different 
$x$, % position due to the local chemical potential, we can 
we obtain information about different dopings in a single calculation. %simulation.
We use DMRG \emph{without} conserving total particle number to allow SC orders to develop. %be finite.
The local densities and the local SC orders along the longitudinal direction are shown in Fig.~\ref{fig:doping_denp}.
The SC orders on the horizontal and vertical bonds are symmetric with opposite signs, showing the $d$-wave symmetry.
The maximum SC order appears around the density $n\approx 0.875$ ($1/8$ hole doping), and decays for both higher and lower densities.
This result is further confirmation of the doping dependence of the SC order observed in Sec.~\ref{ssec:doping-dep}, and  
validates the choice of doping $h\sim 1/8$ as a representative case in studying the SC response.

The optimal doping $\mu_\mathrm{opt}\approx 1.73$ is actually at the boundary between two different stripe fillings.
For $\mu<\mu_\mathrm{opt}$ the ground states are filled stripes ($4$ holes per stripe in the width-4 cylinder) and for $\mu>\mu_\mathrm{opt}$ the ground states are half filled stripes ($2$ holes per stripe).
This can be seen from the more abrupt change in density at $\mu_\mathrm{opt}$ (top panel in Fig.~\ref{fig:doping_denp}), and was further confirmed by 
%the simulations 
our calculations with uniform chemical potentials (not shown here).
This picture is consistent with the idea that fluctuations between different stripe fillings can help induce 
SC orders. 
%picture that the SC can be induced by the fluctuation between different stripe fillings. Since the fluctuation is strongest at the boundary, it induces maximum SC order.

Note that, strictly speaking, the SC order should be zero here in a finite-size system, 
since the Hamiltonian does not break (total) particle number conservation 
without an applied pairing field.
%(no pairing field applied).
However in the DMRG calculation %simulations 
the variational ground states often break the symmetry due to the finite bond dimensions kept.
This feature has been used in the past to study the magnetization and now the SC pairing order.

\begin{figure}[h!]
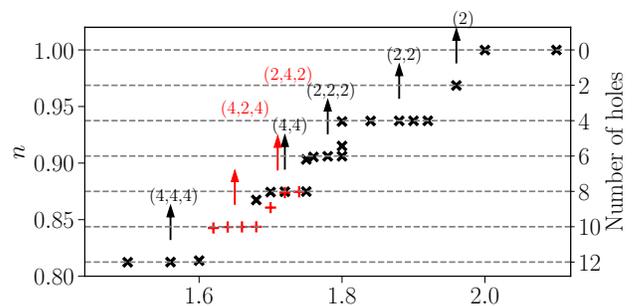
	
\includegraphics[width=0.99\columnwidth]{{{den}}}
\caption{
    The density dependence on $\mu$ for a $16\times 4$ cylinder with pairing field $h_p=0.25$ at the left edge.
    The numbers in the parentheses are the number of holes in each stripe, from the left to the right stripes.
    For example $(4,4)$ means two filled stripes (with $4$ holes).
    The red color indicates the mixture of different stripe fillings.
    }
    \label{fig:doping_denp_16x4}
\end{figure}

In Fig.~\ref{fig:doping_denp_16x4} we show the particle density (number of holes) as a function of chemical potential $\mu$ for a $16\times 4$ cylinder with pinning pairing field $h_p=0.25$ at the left edge.
The density is consistent with the local density in Fig.~\ref{fig:doping_denp}, confirming 
the reliability of the analysis.
The number of holes in each stripe is shown in the parentheses.
Note that the number of holes in a stripe is even for all $\mu$.
This indicates that, although there is no long-range SC order as we conclude in the main text, short-range pairing exists in the stripes.

\section{Filled and $2/3$-filled stripes on 6-leg cylinders}
\label{Sec:3-2_stripe}
\begin{figure}[h!]
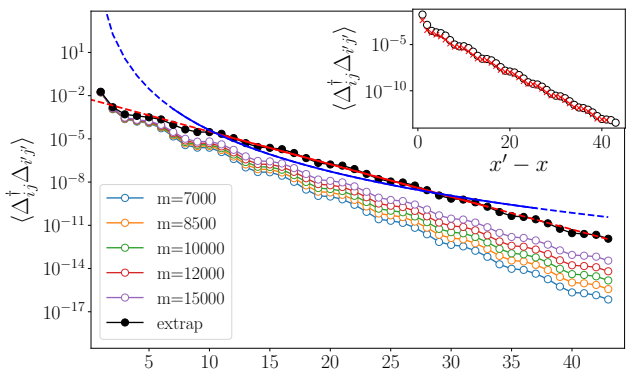
	
	\includegraphics[width=0.99\columnwidth]{{{U8_48x6_4h_pcorr}}}
	\caption{
		%\C{({\em Added correlation length}.)} 
		SC pair-pair correlation for the $2/3$-filled stripes with $1/8$ hole doping on
		$48\times 6$ cylinders. 
        Different curves are for different bond dimensions $m$, as well as extrapolation to infinite bond dimension.
        The pair separation $(x_0 - x)$ is measured with
        respect to reference bond, a vertical bond at $x = 5$. Both
        exponential (red line) and algebraic (blue curve) fits
        are shown; the solid (dash) region indicates the region (not)
        used in the fits.
        The correlation length is $\approx 1.9$ from the fitting.
        The inset shows the (negative) correlations
        on the vertical ($\parallel$) (horizontal ($\perp$)) bonds for $m = 15000$.}
	\label{fig:pair_corr_2_3_stripe}
\end{figure}

Besides the filled stripes, we also considered the $2/3$-filled stripes on 
width-6 cylinders.
The filled and $2/3$-filled stripes are the only two striped states that can be stabilized on 
width-6 cylinders in DMRG.
In Fig.~\ref{fig:pair_corr_2_3_stripe} we show the pair-pair correlations for the $2/3$-filled stripes on a $48\times 6$ system. %cylinders.
The correlations for both the finite bond dimension $m$ as well as the infinite $m$ are shown.
The detail of the extrapolation will be shown in the next section.
Both the power-law and the exponential fittings are shown.
As in the filled stripes, the correlations decay exponentially with distance.
The inset shows the absolute values of the correlations on both the vertical bonds and the horizontal bonds.
The correlations on the horizontal bonds again are perfectly symmetric with the vertical bonds at the same location but
with opposite sign, consistent with the $d$-wave symmetry.

In Fig.~\ref{fig:en} we show the linear extrapolation of the energies with the two-site variance for both the filled and the $2/3$-filled stripes.
The clearly linear behaviors typically indicate the stability of the MPS toward the zero-variance limit.
In other words, the MPS basically stays in the same state for the considered bond dimensions.
The crossing of the lines shows that the filled stripes is lower in energy only when the bond dimension is sufficiently large.
The filled stripes need larger bond dimension than the $2/3$-filled stripes to achieve the same variance, because it contains higher entanglement.

\begin{figure}[t]
	\includegraphics[width=0.9\columnwidth]{{{48x6_en_truncerr}}}
	\caption{Linear extrapolation of ground state energy with the two-site energy variances for the filled and the $2/3$-filled stripes on $48\times 6$ cylinders. The MPS bond dimensions shown for the filled stripes are from $8500$ to $22000$, and for the $2/3$-filled stripes are from $7000$ to $15000$. 
		The extrapolated energy is $-0.7581(6)$ for the filled stripes, and $-0.7574(4)$ for the $2/3$-filled stripes.
        }
	\label{fig:en}
\end{figure}

%------------------------------

\section{Pair-pair correlation extrapolations}

\begin{figure*}[t]
	\includegraphics[width=0.9\columnwidth]{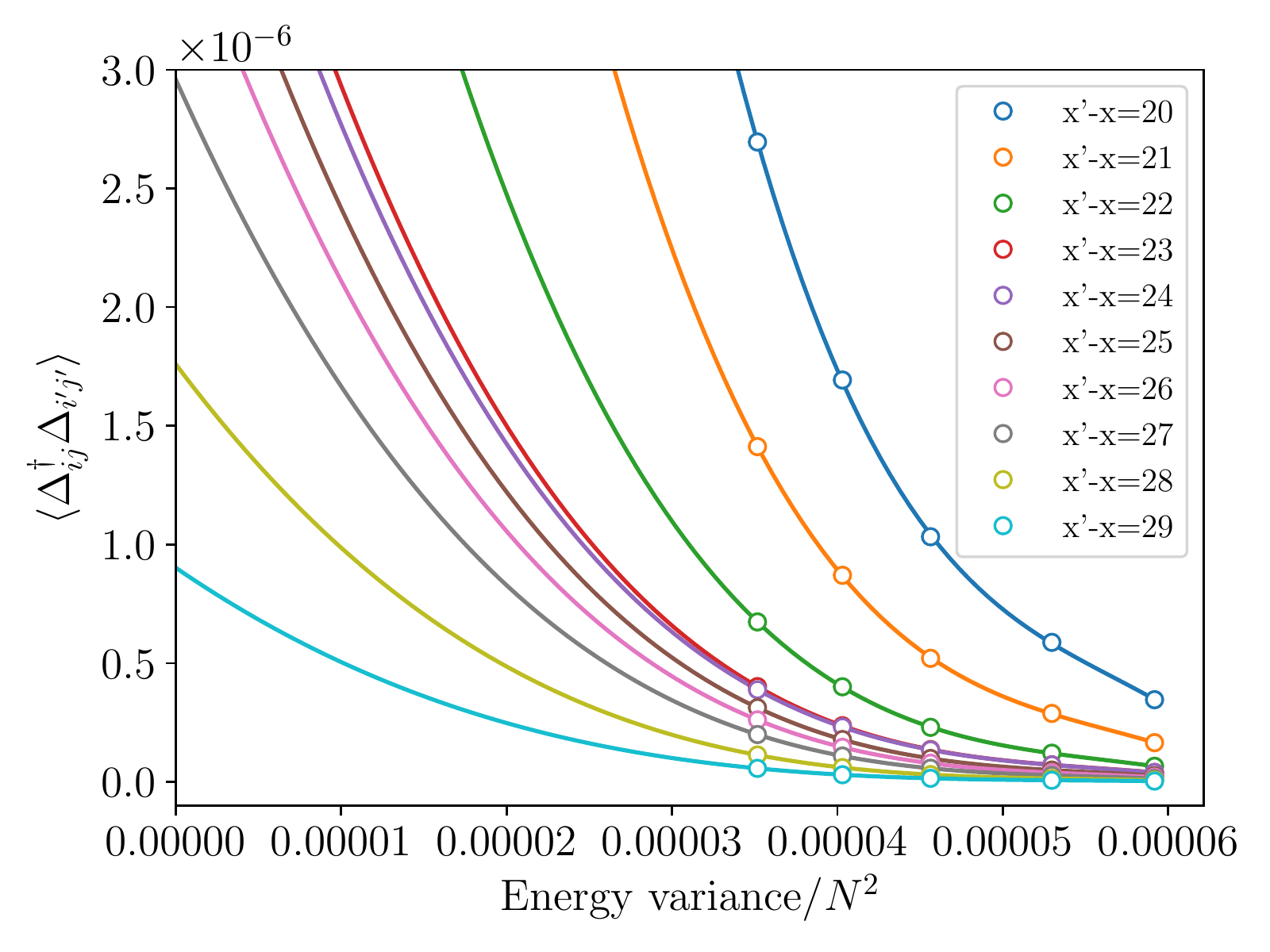}
	\includegraphics[width=0.9\columnwidth]{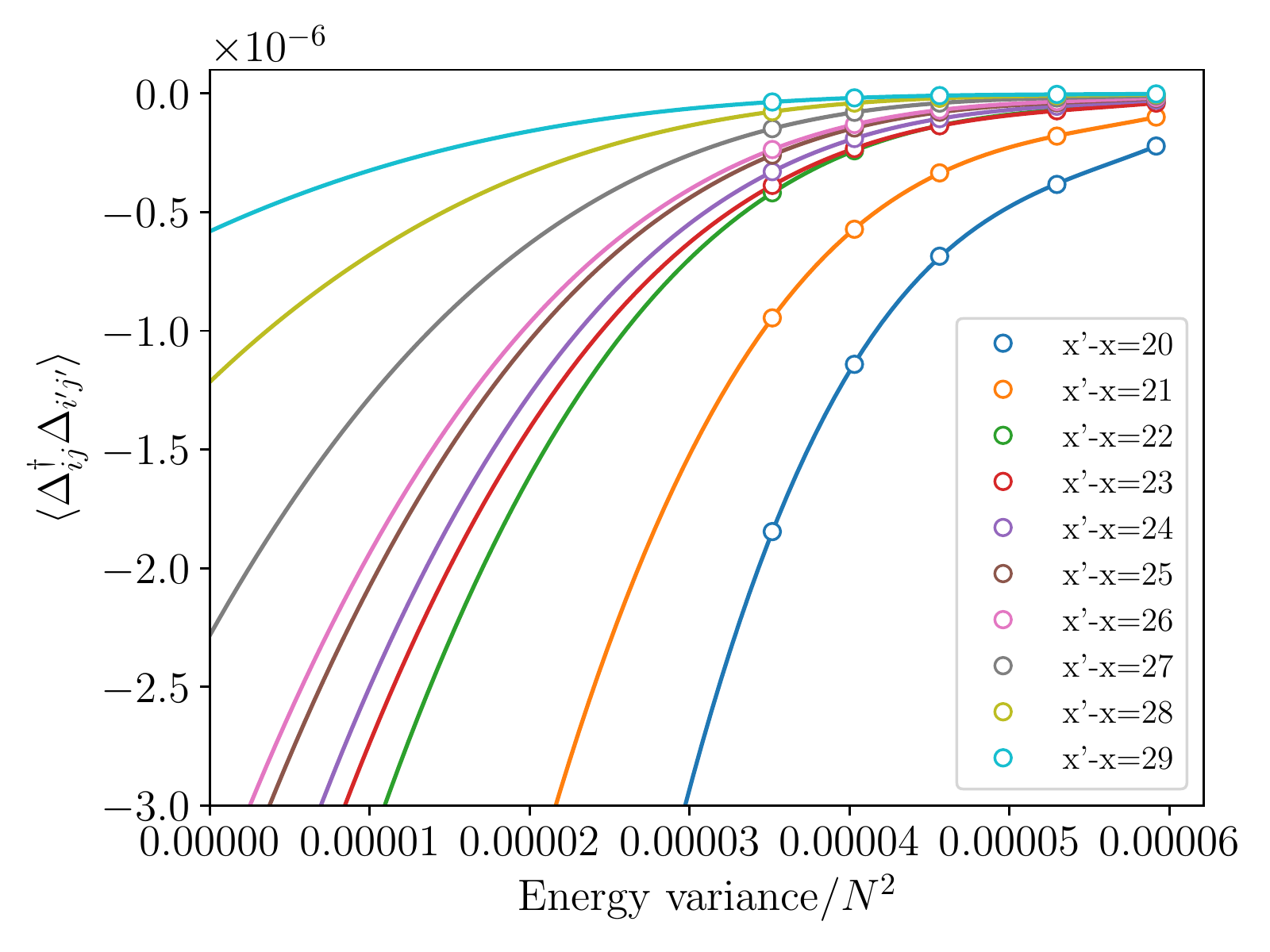}
	\caption{%The cubic extrapolations of 
	Pair-pair correlations (dots) and the cubic extrapolations (curves) by the two-site variance on the (left panel) vertical bonds and the (right panel) horizontal bonds. 
		The reference bond is the vertical bond on $x=5$. 
		The system is the filled stripes on a $48\times 6$ cylinder as in Fig~\ref{fig:pair_corr}.
		The MPS bond dimensions $m$ are from $10000$ to $22000$. }
	\label{fig:pcorrextrap}
\end{figure*} 

\begin{figure*}[t]
	\includegraphics[width=0.9\columnwidth]{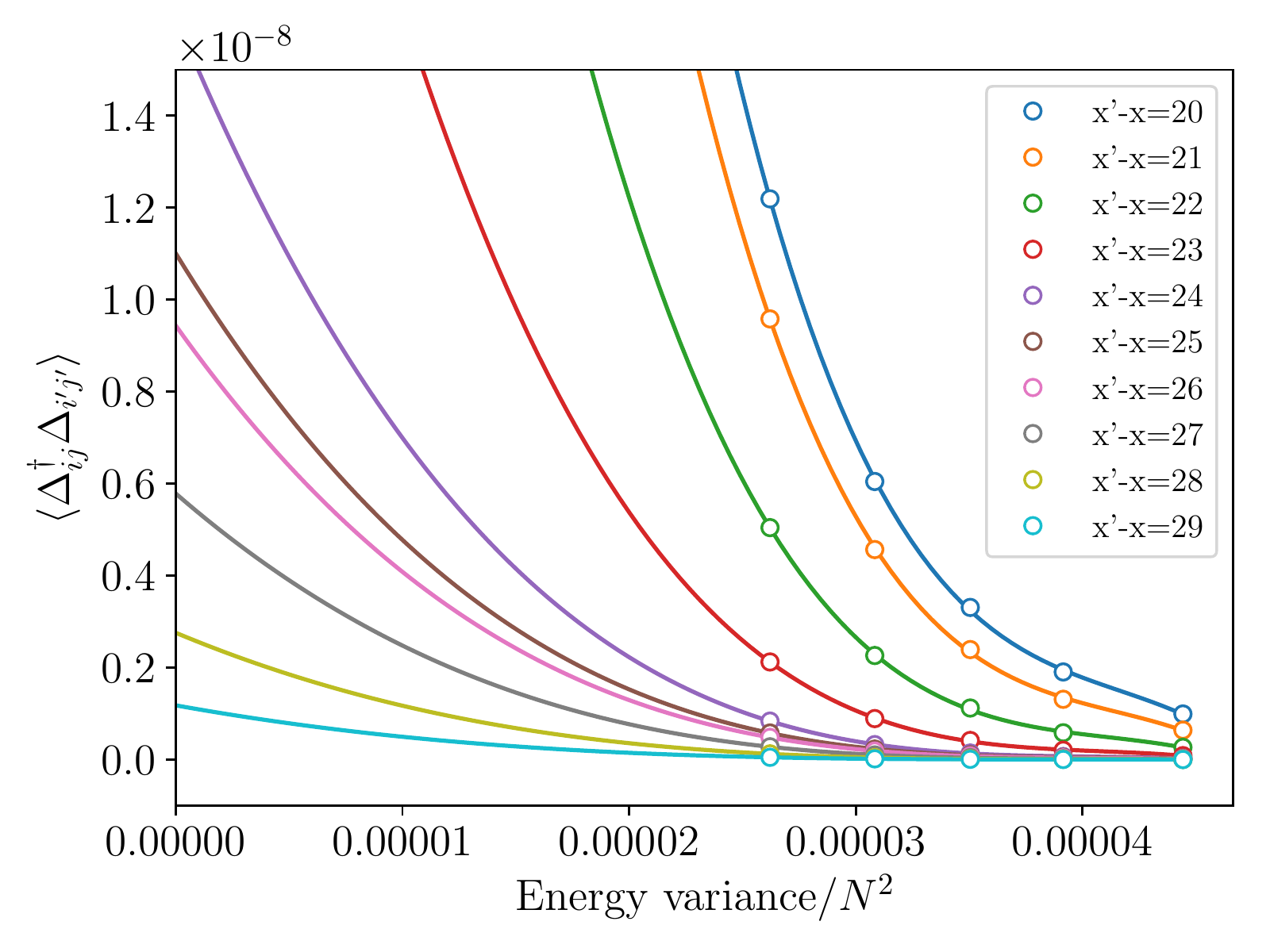}
	\includegraphics[width=0.9\columnwidth]{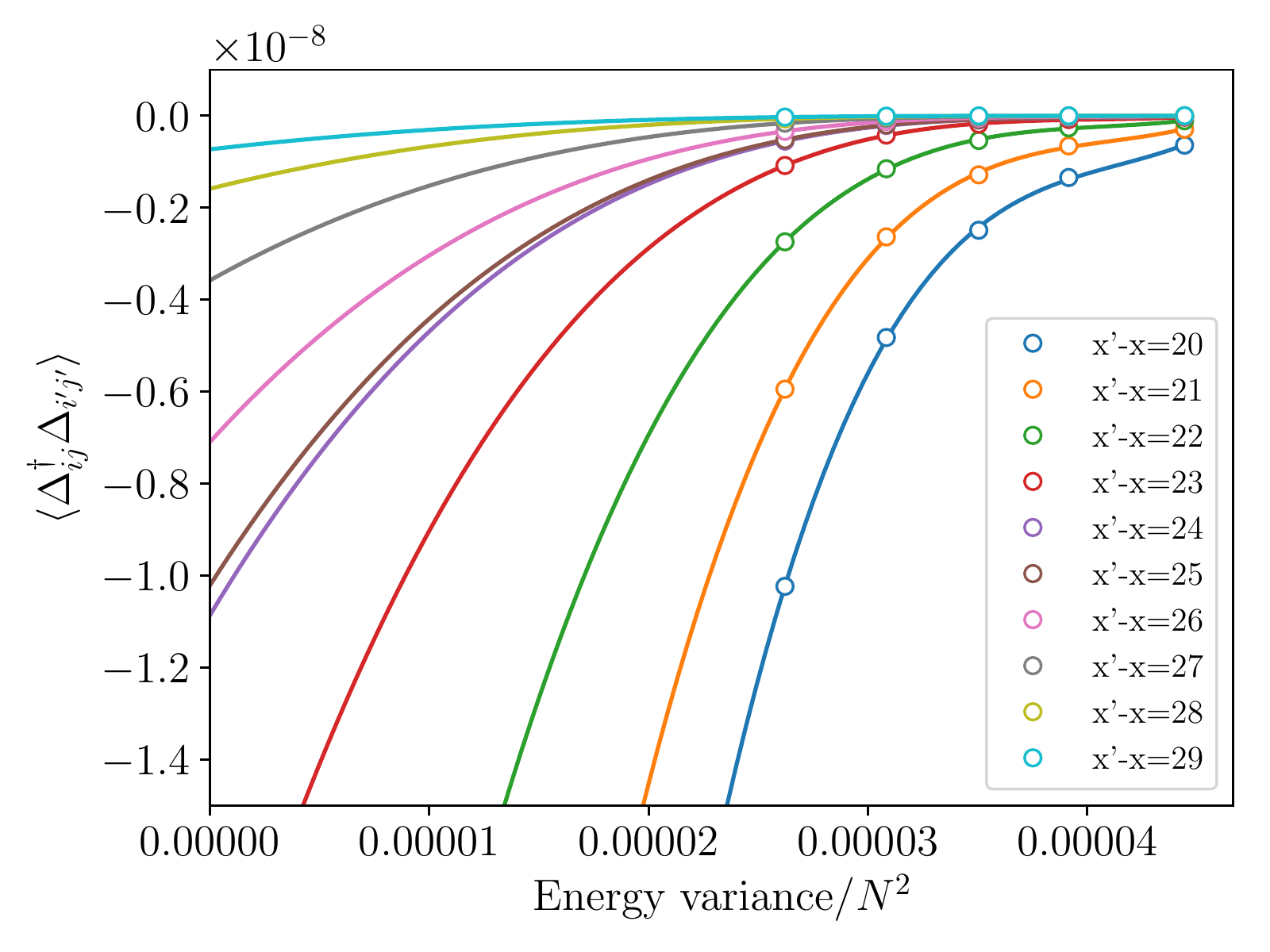}
	\caption{Pair-pair correlations (dots) and the cubic extrapolations (curves) by the two-site variance on the (left panel) vertical bonds and the (right panel) horizontal bonds. 
		The reference bond is the vertical bond on $x=5$. 
		The system is the $2/3$-filled stripes on a $48\times 6$ cylinder as in Fig~\ref{fig:pair_corr_2_3_stripe}.
		The MPS bond dimensions $m$ are from $7000$ to $15000$. }
	\label{fig:pcorrextrap_4h}
\end{figure*} 

Here we discuss some details of the DMRG simulations.
In the simulations preserving $SU(2)$ symmetry, i.e., the simulations on the systems without any pairing field, %the 
temporary local chemical potentials are applied in the first few sweeps on the expected locations of the stripes to stabilize the states and improve %the 
convergence.
One often also applies the magnetic pinning field to stabilize the stripes; however in our cases the magnetic field will break the $SU(2)$ symmetry, so no magnetic pinning field is used.
For the filled and the $2/3$-filled stripes, the temporary chemical potentials are applied up to $m=4000$ ($m=1400$), and then are switched off for the further sweeps of larger $m$.

For the same systems, the single-site update is used in DMRG. To eliminate the finite bond-dimension effect, we employ the extrapolations of physical quantities (for example energy and pair-pair correlations) with the two-site energy variances\cite{PhysRevB.97.045125}.
In two-site DMRG, one usually extrapolates the physical quantities with the so-called truncation error (or alternatively called discarded weight).
However in single-site DMRG, the truncation errors are not well defined.
We thus extrapolate the physical quantities with the two-site energy variance, which is an approximation of the full variance $\langle (\hat{H}-E)^2\rangle/N^2$.
Physically the variance is a perfect quantity to extrapolate with since it measures the distance of the variational state to an eigenstate.
In practice this extrapolation scheme was demonstrated to be as reliable as the extrapolation by the truncation error\cite{PhysRevB.97.045125}.

At the largest bond dimension we can achieve, which to our knowledge is also the largest bond dimension that has been done
to date, the pair-pair correlation vs.~energy variance is not yet reach the linear region.
We thus perform
cubic extrapolations to best fit the data, as shown in Fig.~\ref{fig:pcorrextrap} and Fig.~\ref{fig:pcorrextrap_4h}.
Since the correlations on different distance have quite different scales, we show only for part of the distance which is long enough but away enough from the boundary.
Although here we show only the cubic extrapolations, we also tested %tried 
other extrapolations, for example the linear extrapolation of the last three data points, and the results are very similar to what is presented and lead to the same conclusion (exponential decays).

\end{document}